\documentclass[a4paper,11pt]{article}
\usepackage[T1]{fontenc}
\usepackage{fullpage}
\usepackage{lmodern}
\usepackage{amsmath,amssymb}
\usepackage{float}
\usepackage[font=small]{caption}
\usepackage{titling}
\usepackage{yhmath}
\usepackage{cite}
\usepackage{amsthm}
\usepackage[colorlinks=true,linkcolor=.,citecolor=blue]{hyperref}
\usepackage{latexsym}
\usepackage{indentfirst}
\usepackage{tikz}
\usetikzlibrary{arrows.meta}
\numberwithin{equation}{section}
\allowdisplaybreaks[3]
\def\eps{\epsilon}

\newcommand{\pr}{\partial}
\newcommand{\unc}{{\underline{C}}}
\newcommand{\Ric}{{\ric}}
\newcommand{\g}{\mathbf{g}}
\newcommand{\R}{{\mathbf{R}}}

\def\la{\lambda}
\def\Cb{\unc}
\def\Om{\Omega}
\def\si{\sigma}
\def\b{\beta}
\def\nab{\nabla}
\def\les{\lesssim}
\def\a{\alpha}
\def\b{\beta}
\def\ga{\gamma}
\def\de{\delta}
\def\om{\omega}

\newcommand{\T}{\mathbf{T}}
\def\ric{\mathbf{Ric}}
\newtheorem{thm}{Theorem}[section]
\newtheorem{maintheorem}{Theorem}

\newtheorem{prop}[thm]{Proposition}
\newtheorem{lem}[thm]{Lemma}
\newtheorem{rk}[thm]{Remark}
\title{Sharp decay and regularity thresholds for Minkowski stability in spherical symmetry}
\author{Dawei Shen, Jingbo Wan}
\begin{document}
\maketitle
\begin{abstract}
We establish sharp decay and regularity thresholds for Minkowski stability in the spherically symmetric Einstein--scalar field system. For small smooth characteristic data measured in weighted derivative seminorms, we prove global stability at borderline decay and half-derivative regularity of the initial data. The resulting developments remain uniformly close to Minkowski spacetime and are future causally geodesically complete. Both thresholds are sharp: below either the decay or regularity threshold, we construct arbitrarily small smooth, initially untrapped data whose developments contain trapped spheres.
\end{abstract}
\section{Introduction}\label{sec:introduction}
We determine the decay and regularity thresholds for Minkowski stability in the spherically symmetric Einstein--scalar field system
\begin{equation}\label{eq:esf}
 \Ric_{\a\b}-\frac12 \R \g_{\a\b}=2\T_{\a\b},\qquad\quad\square_\g\phi=0,
\end{equation}
where $\R$ is the scalar curvature and
\[
\T_{\a\b}=\pr_\a\phi\pr_\b\phi-\frac{1}{2}\g_{\a\b}\g^{\ga\de}\pr_\ga\phi\pr_\de\phi.
\]

Without symmetry, Christodoulou--Klainerman~\cite{CK} established global nonlinear stability for small strongly asymptotically flat vacuum data. Lindblad--Rodnianski~\cite{LR} gave an alternative proof in harmonic coordinates, treating both the vacuum and massless scalar field systems. Bieri~\cite{Bieri} weakened the asymptotic decay assumptions and reduced the number of derivatives by one. The first author extended global vacuum stability to arbitrarily slow positive power decay~\cite{ShenMinimal} and established exterior stability at the borderline decay rate~\cite{ShenBorderline}. Luk--Oh--Warnick~\cite{LukOhWarnick} obtained another proof of global vacuum stability under arbitrarily slow positive power decay in a centre-normalized Newman--Unti gauge. Global stability at the critical decay rate remains open for the Einstein vacuum equations without symmetry; see the first author's survey~\cite{ShenSurvey}.

In spherical symmetry, Christodoulou~\cite{ChristodoulouBV} proved global existence for sufficiently small total variation and established a mass concentration criterion for trapped-sphere formation~\cite{Christodoulou1991}. Luk--Oh--Yang~\cite{LOY} proved future causal geodesic completeness under pointwise $C^1$ smallness and sublinear integral growth, allowing infinite BV norms and Bondi mass. Luk--Oh~\cite{LukOh} obtained sharp polynomial decay under local BV scattering and initial decay assumptions. Liu--Zhang~\cite{LiuZhangWave,LiuZhang} proved global existence and completeness under small weighted pointwise $C^1$ bounds with positive power decay.

In this paper, we study the stability of Minkowski spacetime for the characteristic initial value problem of the spherically symmetric Einstein-scalar field. We measure the data using weighted derivative seminorms $\|\cdot\|_{X^{s,q}}$, where $s$ is the decay exponent and $q$ is the regularity exponent (Section~\ref{sec:data-spaces}). As discussed in Section~\ref{sec:scaling-criticality}, this seminorm is critical for large-scale dilation at $s=1$ and for concentration at $q=\tfrac12$.

We first establish a version of the Minkowski stability result at $s=1$ and $q=1$ (Theorem~\ref{thm:s=1}), which is \emph{sharp for the decay threshold}. At this threshold, we give a simple proof using an auxiliary weight $L(v)$ closely related to the borderline weights in the vacuum initial-data construction of Shen--Wan~\cite[Proposition~4.2 and Remark~4.3]{ShenWanIDS}. For nonconstant $\Phi$ with $\int_1^\infty z|\Phi'(z)|^2dz\le\epsilon_0^2$, we obtain $L$ by averaging the initial derivative energy on logarithmic scales:
\[
L(v)^{-2}=\epsilon_0^{-2}\int_1^\infty e^{-|\log(v/z)|/2}z|\Phi'(z)|^2dz,\qquad\int_1^\infty\frac{dv}{vL(v)^2}\leq 4,
\]
where one has $L(v)\to\infty$. While a standard $r^p$--type weight is sufficient to close the bootstrap for $s>1$ (analogous to the approach for the vacuum without symmetry in~\cite{ShenMinimal}), the borderline case $s=1$ requires this auxiliary function $L(v)$ as a replacement to close the bootstrap. When $s>1$ and $q\ge1$, Theorem~\ref{thm:s=1} also shows polynomial convergence of the geometry, similar to the results in~\cite{ShenMinimal} in the vacuum case.

We then generalize this to a result that is \emph{sharp for both the decay and regularity thresholds}. For sufficiently small, smooth, and compatible characteristic data, Theorem~\ref{thm:half-stability} establishes future completeness whenever $s\geq 1$ and $q\geq \tfrac12$: the maximal future development is causally geodesically complete, and the geometry converges uniformly to Minkowski spacetime along outgoing cones. Notably, this requires neither pointwise smallness nor decay of the initial data, and no smallness is imposed on higher derivatives. At the sharp regularity $q=\tfrac12$ and borderline decay $s=1$, we use a scale-invariant energy--bulk estimate. With $\pr_r=\lambda^{-1}\pr_v$ and $\lambda=\pr_v r$, we introduce the energy $E_V(u)$ and bulk $B_V(u)$ as
 \begin{align*}
 E_V(u)&:=[\pr_r(r\phi)(u,\cdot)]_{\dot H^{1/2}(u,V)}^2,\\
 B_V(u)&:=\int_1^u\int_{u'}^V
 \frac{|\pr_r(r\phi)(u',v)-\pr_r(r\phi)(u',u')|^2}{(v-u')^2}dvdu'.
 \end{align*}
We manage to show that
\[
 E_V'(u)+2B_V'(u)\le C E_V(u)B_V'(u),\qquad E_V(u)+B_V(u)\le E_V(1),
\]
where the second inequality follows by absorbing the error for small $E_V$ and integrating in $u$.

Below either threshold, however, Theorem~\ref{thm:instability} provides sequences of smooth, \emph{strictly untrapped} initial data whose developments still form trapped spheres---even though their seminorms tend to zero and their initial geometry converges uniformly to Minkowski. The idea behind this instability result is simply mass concentration for compactly supported initial data, which is untrapped. By exploiting the supercriticality of the norms in this regime, one can rescale such initial data so that it has an arbitrarily small norm while still forming a trapped sphere. To the best of our knowledge, this is the first Minkowski stability result for the Einstein field equations with sharp thresholds in both decay and regularity, even in spherical symmetry.
\subsection{Equations and initial data}\label{sec:intro-equations}
In double-null coordinates,
\begin{equation}\label{eq:geometry-definitions}
\g=-\Om^2dudv+r^2d\om^2,\qquad\la=\pr_vr,\qquad \nu=\pr_ur,\qquad\mu=\frac{2m}{r},\qquad 1-\mu=-4\Om^{-2}\la\nu.
\end{equation}
Here $d\omega^2$ is the unit round metric. We denote the outgoing and incoming null curves, respectively, as
\[
C_u:=\{(u,v'):v'\ge u\},\qquad\Cb_v:=\{(u',v):1\le u'\le v\}.
\]
The axis is denoted by
\[
\Gamma:=\{(u,u):u\geq 1\}.
\]
On an untrapped region, \eqref{eq:esf} under spherical symmetry reduces to
\begin{equation*}\tag{SSESF}\label{eq:ssesf}
 \begin{aligned}
 \pr_u\pr_vr&=\frac{2m\nu\lambda}{(1-\mu)r^2},&\pr_u\pr_v(r\phi)&=\frac{2m\nu\lambda}{(1-\mu)r^2}\phi,\\
 \nu\pr_um&=\frac12(1-\mu)r^2(\pr_u\phi)^2,&\lambda\pr_vm&=\frac12(1-\mu)r^2(\pr_v\phi)^2.
 \end{aligned}
\end{equation*}
Prescribe $\Phi$ on $C_1$, with regular vertex $(1,1)$. Fix $v$ by $\lambda=1/2$ on $C_1$ and fix $u$ by $u=v$ on the axis. Moreover, we impose
\begin{equation}\label{eq:data}
\begin{aligned}
u=1,\ v\ge1:&\qquad\lambda(1,v)=\frac12,\qquad 2\pr_v(r\phi)(1,v)=\Phi(v),\\
u=v:&\qquad r(u,u)=m(u,u)=0.
\end{aligned}
\end{equation}
Thus $r(1,v)=(v-1)/2$ and
\[
(\lambda^{-1}\pr_v)(r\phi)(1,v)=\Phi(v).
\]
The constraints determine the remaining initial geometry on $C_1$.
\subsection{The data spaces}\label{sec:data-spaces}
For $s\in\mathbb R$ and $q\ge0$, define
\[
X^{s,q}:=\left\{\Phi\in L^2_{\mathrm{loc}}([1,\infty)):\|\Phi\|_{X^{s,q}}<\infty\right\},
\]
where, for integer $q\ge0$,
\begin{equation}\label{eq:X-norm}
\|\Phi\|_{X^{s,q}}:=\left(\sum_{j=1}^q\int_1^\infty v^{s-1}|(v\pr_v)^j\Phi(v)|^2\frac{dv}{v}\right)^{1/2}.
\end{equation}
For noninteger $q=k+\beta$, where $k\ge0$ is an integer and $0<\beta<1$, set
\begin{equation}\label{eq:X-fractional}
\begin{aligned}
&\|\Phi\|_{X^{s,k+\beta}}:=\Bigg(\sum_{j=1}^k\int_1^\infty v^{s-1}|(v\pr_v)^j\Phi(v)|^2\frac{dv}{v}\\
 &+(1-\beta)\int_1^\infty\int_1^\infty
 \frac{\min\{v,v'\}^{s-1}(vv')^{\beta-1/2}}{|v-v'|^{1+2\beta}}
 \left|(v\pr_v)^k\Phi(v)-(v'\pr_{v'})^k\Phi(v')\right|^2dvdv'\Bigg)^{1/2}.
\end{aligned}
\end{equation}
Adding a constant to $\Phi$ adds the same constant to $\phi$ and leaves the geometry unchanged. In particular,
\begin{equation}\label{eq:X-inclusions}
\begin{split}
X^{s,0}&=L^2_{\mathrm{loc}}([1,\infty)),\qquad\qquad
 \|\Phi\|_{X^{s,0}}=0,\\
\|\Phi\|_{X^{s,1}}&=\left(\int_1^\infty v^s|\Phi'(v)|^2dv\right)^{1/2},\\
\|\Phi\|_{X^{1,1}}&\le\|\Phi\|_{X^{s,q}},\qquad s\ge1,\quad q\ge1.
\end{split}
\end{equation}
For $s=1$ and $q=1/2$, the definition reduces to
\begin{equation}\label{half:eq:data-norm}
\|\Phi\|_{X^{1,1/2}}=\left(\frac12\int_1^\infty\int_1^\infty\frac{|\Phi(v)-\Phi(v')|^2}{|v-v'|^2}dvdv'\right)^{1/2}.
\end{equation}
All existence statements below concern smooth data compatible with \eqref{eq:data}; no smallness of higher derivatives is assumed.
\subsection{Main results}\label{sec:main-results}
We begin with stability for small characteristic data in $X^{s,1}$, with $s\geq 1$, including the borderline decay case of $s=1$.
\begin{maintheorem}\label{thm:s=1}
Let $s\geq 1$. Let $\Phi$ be smooth and compatible with a regular vertex. Assume that
\begin{equation}\label{eq:small}
\|\Phi\|_{X^{s,1}}\le\eps_0.
\end{equation}
For sufficiently small $\eps_0$, the characteristic problem \eqref{eq:ssesf}--\eqref{eq:data} admits a unique smooth regular solution on $1\leq u\leq v<\infty$. Its development is future causally geodesically complete and satisfies the following estimates:
\begin{equation}\label{eq:geometry-main}
 \sup_{1\leq u\leq v}\left(\left|\la-\frac12\right|+\left|\nu+\frac12\right|+|\Om^2-1|+\mu+|r\pr_v\phi|\right)\les\eps_0.
\end{equation}
Moreover, the scalar field satisfies
\begin{equation}\label{eq:one-derivative-energy}
\sup_{u\geq 1}\int_u^\infty r\left|(\la^{-1}\pr_v)^2(r\phi)(u,v)\right|^2\la\,dv\les\eps_0^2
\end{equation}
and
\begin{equation}\label{eq:energy-limit}
\int_u^\infty r\left|(\la^{-1}\pr_v)^2(r\phi)(u,v)\right|^2\la\,dv\to0\qquad\text{as }u\to\infty.
\end{equation}
The geometry approaches Minkowski spacetime uniformly along the outgoing cones:
\begin{equation}\label{eq:geometry-limit}
\sup_{v>u}\left(\left|\la-\frac12\right|+\left|\nu+\frac12\right|+|\Om^2-1|+\mu+\left|\frac{2r}{v-u}-1\right|+|r\pr_v\phi|\right)\to0\qquad\text{as }u\to\infty.
\end{equation}
For $s>1$, the expression on the left-hand side of~\eqref{eq:geometry-limit} is $O(\eps_0u^{-\min\{(s-1)/2,1/4\}})$.
\end{maintheorem}
The next theorem lowers the regularity from $q=1$ to $q=\tfrac12$ at the critical decay $s=1$.
\begin{maintheorem}
\label{thm:half-stability}
Suppose that $\Phi$ is smooth, compatible with a smooth regular vertex, and
\[
\|\Phi\|_{X^{1,1/2}}\le\eps_0.
\]
For sufficiently small $\eps_0>0$, the characteristic problem with normalization \eqref{eq:data} admits a unique smooth regular solution on $1\le u\le v<\infty$, whose development is future causally geodesically complete and satisfies
\begin{equation}\label{eq:geometry-main-12}
\sup_{1\le u\le v}\left(\left|\la-\frac12\right|+\left|\nu+\frac12\right|+\mu+|\Om^2-1|\right)\les\eps_0.
\end{equation}
Moreover, the scalar field satisfies
\begin{equation}\label{half:eq:main-energy}
\sup_{u\ge1}[\la^{-1}\pr_v(r\phi)(u,\cdot)]_{\dot H^{1/2}(u,\infty)}^2+\int_1^\infty\int_u^\infty\frac{|\la^{-1}\pr_v(r\phi)(u,v)-\la^{-1}\pr_v(r\phi)(u,u)|^2}{(v-u)^2}dvdu\les\eps_0^2,
\end{equation}
and
\begin{equation}\label{half:eq:energy-decay}
[\la^{-1}\pr_v(r\phi)(u,\cdot)]_{\dot H^{1/2}(u,\infty)}\to0\qquad\text{as }u\to\infty.
\end{equation}
The geometry converges uniformly to Minkowski spacetime along the outgoing cones:
\begin{equation}\label{eq:geometry-limit-12}
\sup_{v>u}\left(\left|\la-\frac12\right|+\left|\nu+\frac12\right|+\mu+|\Omega^2-1|+\left|\frac{2r}{v-u}-1\right|\right)\to0\qquad\text{as }u\to\infty.
\end{equation}
The same conclusions hold for sufficiently small smooth compatible data in $X^{s,q}$, for every fixed $s\ge1$ and $q\ge\tfrac12$.
\end{maintheorem}
We note that the smallness condition in Theorem~\ref{thm:half-stability} is incomparable with small total variation and the criterion of~\cite{LOY}, which can be seen by three explicit examples of data.
\begin{enumerate}
\item[(i)] For $0<a\le\eps_0$, set
\begin{equation}\label{Phiosc}
\Phi(v)=\sin\bigl(a\log(1+\log v)\bigr).
\end{equation}
Then $\|\Phi\|_{X^{1,1}}^2\le a^2\le\eps_0^2$, but the total variation is infinite, $\int_1^\infty|\Phi'(v)|d v=\infty$. Moreover, $\limsup_{v\to\infty}\frac1v\int_v^{2v}|\Phi(z)|d z=1$, so $\Phi$ fails the sublinear integral bound of~\cite[Theorem~1.1]{LOY}.
\item[(ii)] For $a>0$ and $0<\de<1/2$, choose smooth $\Phi$ supported in $[2,4]$, increasing from $0$ to $a$ on $[2,2+\de]$, equal to $a$ on $[2+\delta,3]$, and decreasing to $0$ on $[3,4]$. Then $\int_1^\infty|\Phi'|dv=2a$ and $\|\Phi\|_{X^{1,1/2}}^2\ge a^2\log\frac{1+\delta}{2\delta}$. If we take $\delta=e^{-a^{-4}}$, the total variation tends to zero and the critical seminorm tends to infinity as $a\downarrow0$.

\item[(iii)] Let $0\le\chi\in C_c^\infty((1,6))$ equal $1$ on $[3,4]$. The profiles $\Phi(v)=a\sum_{k=3}^\infty\chi(v-1-2^k)$ in~\cite[proof of Corollary~1.4]{LOY} satisfy, for $v\ge u\ge1$,
\[
 |\Phi(v)|+|\Phi'(v)|\les_\chi a,\qquad
 \int_u^v|\Phi(z)|dz\les_\chi a(v-u)^{1/2},
\]
but the critical seminorm is not finite
\[
 \|\Phi\|_{X^{1,1/2}}^2
 \ge a^2\sum_{k=3}^\infty
 \int_{2^k+4}^{2^k+5}\int_{2^k+7}^{2^k+8}
 \frac{dv'dv}{(v'-v)^2}
 =a^2\log\frac98\sum_{k=3}^\infty1=\infty.
\]
\end{enumerate}

\begin{figure}[H]
\centering
\begingroup
\definecolor{initialblue}{RGB}{35,88,125}
\definecolor{trappedred}{RGB}{153,48,42}
\tikzset{
 causal edge/.style={line width=.65pt},
 infinity edge/.style={densely dotted,line width=.7pt},
 initial edge/.style={draw=initialblue,line width=1.15pt},
 transfer edge/.style={draw=initialblue,dashed,line width=.8pt},
 null direction/.style={draw=black!45,line width=.45pt,-{Stealth[length=3.5pt]}},
 causal label/.style={font=\footnotesize,inner sep=1.5pt},
 sphere/.style={circle,fill,inner sep=1.5pt},
 ideal point/.style={circle,draw,line width=.65pt,fill=white,inner sep=1.7pt}}
{\footnotesize
$C_1=\{u=1\}:\qquad \lambda=\tfrac12,\qquad 2\pr_v(r\phi)=\Phi.$
\par}\medskip
\begin{minipage}[t]{.45\textwidth}
\centering\vspace{0pt}
{\color{initialblue}\rule{0pt}{14pt}\strut$s\ge1,\quad q\ge\tfrac12$}\par\smallskip
\begin{tikzpicture}[x=1.1cm,y=1.1cm]
 \path[use as bounding box] (-.75,-.4) rectangle (3.1,4.6);
 \fill[black!3] (0,0)--(2.1,2.1)--(0,4.2)--cycle;
 \draw[causal edge] (0,0)--(0,4.2);
 \draw[infinity edge] (0,4.2)--(2.1,2.1);
 \draw[initial edge] (0,0)--(2.1,2.1)
   node[pos=.51,below,sloped,causal label,text=initialblue]
        {$C_1,\ \Phi$};
 \node[causal label,rotate=90] at (-.25,1.9)
       {$u=v,\quad r=m=0$};
 \node[causal label,rotate=-45,anchor=south] at (1.12,3.08)
       {$\mathcal I^+$};
 \node[ideal point] at (0,4.2) {};
 \node[causal label,anchor=south] at (0,4.3) {$i^+$};
 \draw[null direction] (.3,.95)--(.8,1.45);
 \draw[null direction] (1.3,1.8)--(.8,2.3);
 \node[causal label] at (.58,2.78) {$\mu<1$};
 \fill (0,0) circle (1.25pt);
 \node[causal label,anchor=north] at (0,-.08) {$(1,1)$};
\end{tikzpicture}\par
\small $\|\Phi\|_{X^{s,q}}\le\eps_0$\par
Future causal geodesic completeness\par
\footnotesize $\mu\to0\quad(u\to\infty)$\par
Theorems~\ref{thm:s=1}--\ref{thm:half-stability}
\end{minipage}
\hfill
\begin{minipage}[t]{.45\textwidth}
\centering\vspace{0pt}
{\color{trappedred}\rule{0pt}{14pt}\strut$s<1\ \text{or}\ q<\tfrac12$}\par\smallskip
\begin{tikzpicture}[x=1.1cm,y=1.1cm]
 \path[use as bounding box] (-.75,-.4) rectangle (3.1,4.6);
 \draw[causal edge,-{Stealth[length=3.5pt]}] (0,0)--(0,1.6);
 \node[causal label,rotate=90,align=center] at (-.35,.8)
       {$u=v$\\$r=m=0$};
 \draw[initial edge,-{Stealth[length=4pt]}] (0,0)--(2.55,2.55)
   node[pos=.46,above,sloped,causal label,text=initialblue]
        {$\lambda>0,\ \nu<0$}
   node[pos=.42,below,sloped,causal label,text=initialblue]
        {$C_1,\ \Phi_N$};
 \node[sphere,fill=trappedred] at (.95,2.7) {};
 \node[causal label,text=trappedred,anchor=west,align=left]
       at (1.11,2.67) {trapped\\$\mu>1$};
 \draw[draw=trappedred,line width=.8pt,-{Stealth[length=4pt]}]
        (.95,2.7)--(1.58,3.33);
 \draw[draw=trappedred,line width=.8pt]
        (1.58,3.33)--(2.45,4.2);
 \draw[draw=trappedred,line width=.8pt]
       (2.36,4.11)--(2.54,4.29) (2.36,4.29)--(2.54,4.11);
 \node[causal label,align=center,text=trappedred] at (.72,3.6)
       {finite affine\\length};
 \fill (0,0) circle (1.25pt);
 \node[causal label,anchor=north] at (0,-.08) {$(1,1)$};
\end{tikzpicture}\par
\small $\|\Phi_N\|_{X^{s,q}}\to0$\par
Trapped sphere formation\par
Future null incompleteness\par
\footnotesize Theorem~\ref{thm:instability}
\end{minipage}
\endgroup
\caption{Left: all sufficiently small smooth compatible data at and above both thresholds; right: collapsing sequences below either threshold.}
\label{fig:characteristic-thresholds}
\end{figure}
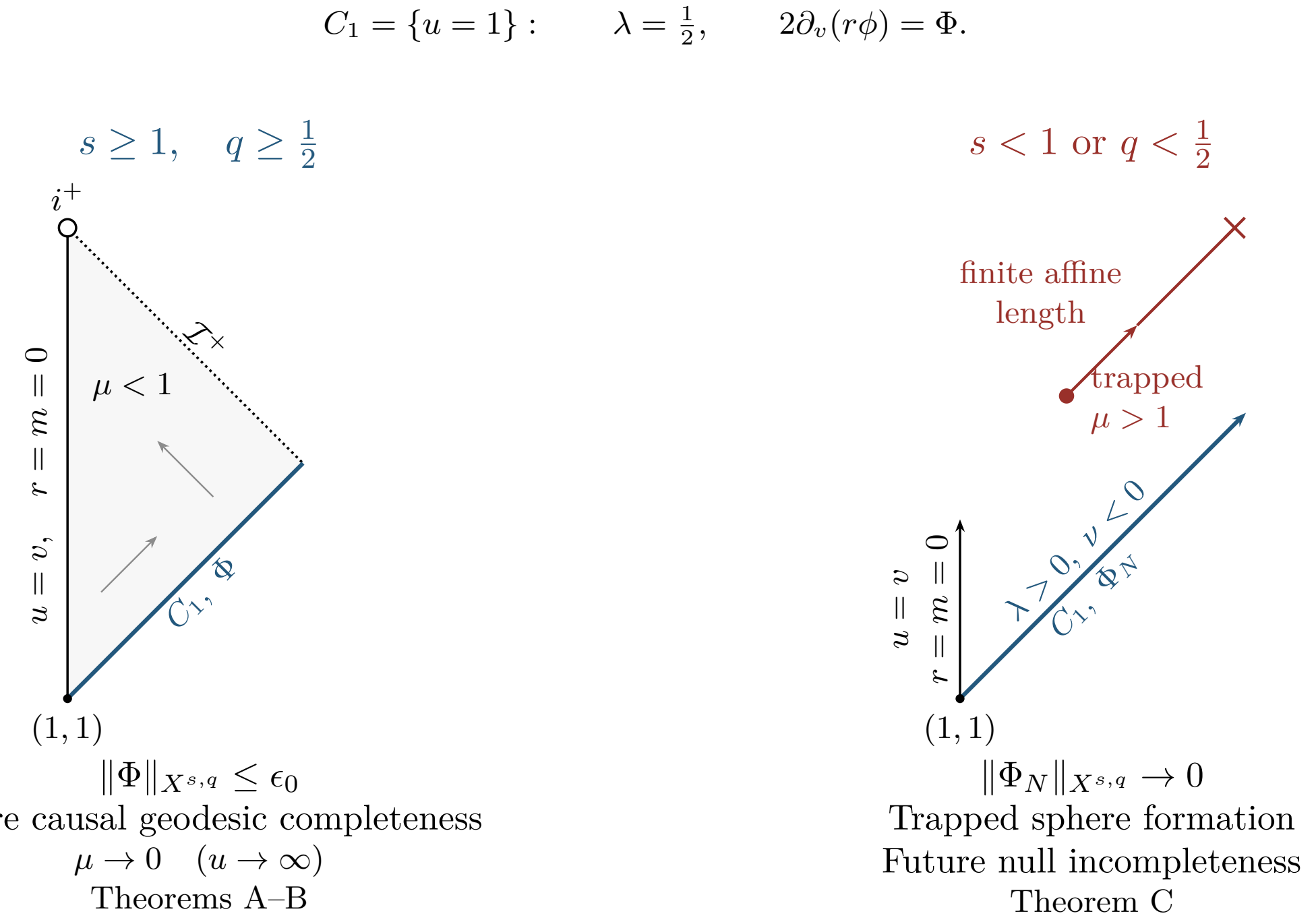

Next, we show that both thresholds $s=1$ and $q=1/2$ are indeed sharp.
\begin{maintheorem}\label{thm:instability}\label{thm:half-sharp}
There exist sequences of smooth data $\Phi_N\in C_c^\infty((1,\infty))$ with the following properties.
\begin{enumerate}
\item[(i)] For each $s<1$, a sequence satisfies
\begin{equation}\label{eq:small-data-sequence}
\|\Phi_N\|_{X^{s,q}}\to0\qquad\text{for every fixed real }q\geq 0.
\end{equation}
\item[(ii)] There is a sequence supported in $[3/2,3]$ such that
\[
\|\Phi_N\|_{X^{s,q}}\to0\qquad\text{for every fixed }s\in\mathbb R\text{ and }0\le q<\frac12.
\]
\end{enumerate}
In both cases, the initial geometry satisfies
\begin{equation}\label{eq:small-geometry-sequence}
 \sup_{v\ge1}\left(
 \left|\lambda_N-\frac12\right|+\left|\nu_N+\frac12\right|
 +|\Omega_N^2-1|+\mu_N\right)(1,v)\to0.
\end{equation}
Each initial cone is smooth and everywhere strictly untrapped:
\begin{equation}\label{eq:initial-untrapped-sequence}
\begin{split}
\la_N(1,v)&=\frac12,\qquad\qquad \nu_N(1,v)\le-\frac12,\\
0\le\mu_N(1,v)&\le CNe^{-N}\le\frac12\qquad(v>1).
\end{split}
\end{equation}
Every maximal development contains a strictly trapped sphere at $u>1$, where $\la_N<0$, $\nu_N<0$, and $\mu_N>1$. Each maximal smooth future development is future null geodesically incomplete.
\end{maintheorem}
Figure~\ref{fig:characteristic-thresholds} shows the corresponding causal diagrams for Theorems~\ref{thm:s=1}, \ref{thm:half-stability} and \ref{thm:instability}.
Set $x=1/r$ and attach $\mathcal I^+=\{x=0,\ u\ge1\}$ in the smooth product coordinates $(u,x,\omega)$, with a smooth angular atlas on $\mathbb S^2$. Here $C^k$ regularity is taken in these coordinates up to $x=0$, locally for $1\le u<\infty$. The Bondi mass is $M_B(u):=\lim_{v\to\infty}m(u,v)\in[0,\infty]$. For $c\in\mathbb R$, the radiation field relative to $c$ is $\lim_{v\to\infty}r(\phi-c)(u,v)$ when this limit exists and is finite.
\begin{maintheorem}\label{thm:null-infinity}
Let $\Phi$ be smooth and compatible with a smooth regular vertex. If $\|\Phi\|_{X^{1,1/2}}\le\eps_0$ for sufficiently small $\eps_0$, its characteristic development satisfies the following.
\begin{enumerate}
\item[(i)] The metric $x^2\g$ has a $C^0$ Lorentzian extension with null boundary $\mathcal I^+$. Every point lies on an outgoing null ray reaching $\mathcal I^+$ at infinite physical affine length. The limit $\nu_\infty(u):=\lim_{v\to\infty}\nu(u,v)<0$ is continuous; set $U(u):=-2\int_1^u\nu_\infty(u')du'$, with $U'\asymp1$.
\item[(ii)] If $\|\Phi\|_{X^{s,q}}<\infty$, $s>1$, $q\ge1$, then $\Phi_\infty:=\lim_{v\to\infty}\Phi(v)$ exists and $\phi(u,v)\to\Phi_\infty$. For $s\ge2$, $M_B(u)<\infty$ and $M_B(u)\downarrow0$; $M_B\in C^1$ and the following limit exists:
\begin{equation}\label{null:eq:mass-loss}
 -\frac{dM_B}{dU}
 =\frac1{4\nu_\infty(u)^2}
 \left|\lim_{v\to\infty}r\pr_u\phi(u,v)\right|^2,
\end{equation}
\begin{equation}\label{null:eq:total-flux}
 \int_1^\infty\frac1{-2\nu_\infty(u)}
 \left|\lim_{v\to\infty}r\pr_u\phi(u,v)\right|^2du=M_B(1).
\end{equation}
Here $M_B(1)=\lim_{v\to\infty}m(1,v)$ is the total initial Bondi mass on $C_1$. For $s>3$, the radiation field $\lim_{v\to\infty}r(\phi-\Phi_\infty)$ relative to $\Phi_\infty$ exists and is $C^1$ in $u$. All $v\to\infty$ limits are locally uniform in $u$.
\item[(iii)] In \textup{(ii)}, for $1<s<2$, $\phi$ is $C^{0,(s-1)/2}$ and $x^2\g$ is $C^{0,s-1}$ in $x$, locally uniformly in $u$. The metric exponent is sharp in $X^{s,1}$. For $s\ge2$, $x^2\g$ is $C^1$ up to $\mathcal I^+$.
\item[(iv)] The decay thresholds in \textup{(ii)}--\textup{(iii)} are sharp: smooth data arbitrarily small in the indicated seminorm can have
\begin{itemize}
\item[\textup{(a)}] in $X^{1,1}$, no limit of $\phi(u,v)$ as $v\to\infty$ on every $C_u$;
\item[\textup{(b)}] in $X^{s,1}$ for each $1\le s<2$, $M_B(u)=\infty$ on every $C_u$ and $x^2\g\notin C^1$ at the initial cut;
\item[\textup{(c)}] in $X^{3,1}$, $\Phi_\infty=0$, finite $M_B(u)\downarrow0$ but $r\phi(u,v)\to\infty$ on every $C_u$.
\end{itemize}
\end{enumerate}
\end{maintheorem}

\begin{figure}[H]
\centering
\begingroup
\definecolor{stabilityblue}{RGB}{35,88,125}
\definecolor{instabilityred}{RGB}{153,48,42}
\begin{tikzpicture}[x=2.5cm,y=3.04175cm,
 region label/.style={font=\footnotesize,align=center,inner sep=2pt},
 tick label/.style={font=\footnotesize,inner sep=1.5pt},
 boundary label/.style={font=\footnotesize,align=center,anchor=north,inner sep=1pt}]
 \fill[instabilityred!9] (-1.15,0) rectangle (4.12,1.75);
 \fill[stabilityblue!9] (1,.5) rectangle (4.12,1.75);
 \fill[stabilityblue!19] (1,1) rectangle (4.12,1.75);
 \draw[stabilityblue!60,densely dotted,line width=.6pt]
       (1,1)--(4.12,1);
 \draw[gray,dashed,line width=.6pt] (-1.15,.5)--(1,.5);
 \draw[gray,dashed,line width=.6pt] (1,0)--(1,.5);
 \draw[stabilityblue,line width=1pt] (1,.5)--(1,1.75);
 \draw[stabilityblue,line width=1pt,-{Stealth[length=3.5pt]}]
       (1,.5)--(4.2,.5);
 \fill[stabilityblue] (1,.5) circle (1.6pt);
 \node[tick label,anchor=south east,fill=white,fill opacity=.85,
       text opacity=1] at (.98,.53) {$(1,\tfrac12)$};
 \draw[line width=.65pt,{Stealth[length=3.5pt]}-{Stealth[length=3.5pt]}]
       (-1.28,0)--(4.22,0);
 \draw[line width=.65pt,-{Stealth[length=3.5pt]}]
       (-1.15,0)--(-1.15,1.96);
 \node[tick label,anchor=south] at (-1.15,1.99) {$q$};
 \node[tick label,anchor=west] at (-1.11,1.85) {$+\infty$};
 \node[tick label,anchor=north] at (-1.26,-.035) {$-\infty$};
 \node[tick label,anchor=south east] at (4.1,.025) {$+\infty$};
 \node[tick label,anchor=north west] at (4.3,-.015) {$s$};
 \foreach \x in {0,1,2,3}{
   \draw[line width=.55pt] (\x,-.02)--(\x,.02);
   \node[tick label,anchor=north] at (\x,-.035) {$\x$};
 }
 \node[tick label,anchor=east] at (-1.18,.05) {$0$};
 \foreach \y/\lab in {.5/{\tfrac12},1/1}{
   \draw[line width=.55pt] (-1.18,\y)--(-1.12,\y);
   \node[tick label,anchor=east] at (-1.18,\y) {$\lab$};
 }
 \node[region label,text=instabilityred,text width=4.7cm]
 at (-.1,1.14) {\textbf{Trapped-sphere examples}\\
       Future null incompleteness\\Theorem~\ref{thm:instability}};
 \node[region label,text=stabilityblue,text width=7.2cm]
 at (2.55,1.37) {\textbf{Stability and polynomial convergence}\\
       $s>1,\quad q\ge1$\qquad Theorem~\ref{thm:s=1}};
 \node[region label,text=stabilityblue,text width=7.2cm,inner sep=1pt]
 at (2.55,.75) {\textbf{Stability and causal completeness}\\
       $C^0$ null boundary\qquad Theorems~\ref{thm:half-stability},~\ref{thm:null-infinity}};
 \node[region label,text=instabilityred,text width=12.8cm,inner sep=1pt]
 at (1.45,.24) {\textbf{Trapped-sphere examples}\qquad
       $0\le q<\tfrac12$\qquad Theorem~\ref{thm:instability}};
 \draw[gray!55,line width=.4pt] (1,-.25)--(4.12,-.25);
 \foreach \x in {1,2,3}{
   \draw[gray!55,line width=.4pt] (\x,-.21)--(\x,-.3);
 }
 \node[boundary label,text=stabilityblue,text width=4.5cm]
 at (-.15,-.32) {Theorem~\ref{thm:null-infinity}\\
       $\|\Phi\|_{X^{1,1/2}}\ll1$\\
       $q\ge1,\quad\|\Phi\|_{X^{s,q}}<\infty$};
 \node[boundary label,text=stabilityblue,text width=2.35cm]
 at (1.5,-.32) {$1<s<2$\\
       $\phi\in C^{0,(s-1)/2}$\\$x^2\g\in C^{0,s-1}$};
 \node[boundary label,text=stabilityblue,text width=2.35cm]
 at (2.5,-.32) {$s\ge2$\\$x^2\g\in C^1$\\finite $M_B$\\finite scalar flux};
 \node[boundary label,text=stabilityblue,text width=2.8cm]
 at (3.6,-.32) {$s>3$\\finite radiation field\\$C^1$ in $u$};
\end{tikzpicture}
\endgroup
\caption{Blue: stability for all sufficiently small smooth compatible data; red: existence of arbitrarily small initially untrapped data forming trapped spheres. Solid boundaries belong to the stable region. The lower labels record the conclusions of Theorem~\ref{thm:null-infinity}, with H\"older regularity measured in $x=1/r$. For $s\geq 2$, the total scalar radiation flux through $\mathcal I^+$ is $M_B(1)$.}
\label{fig:parameter-regions}
\end{figure}
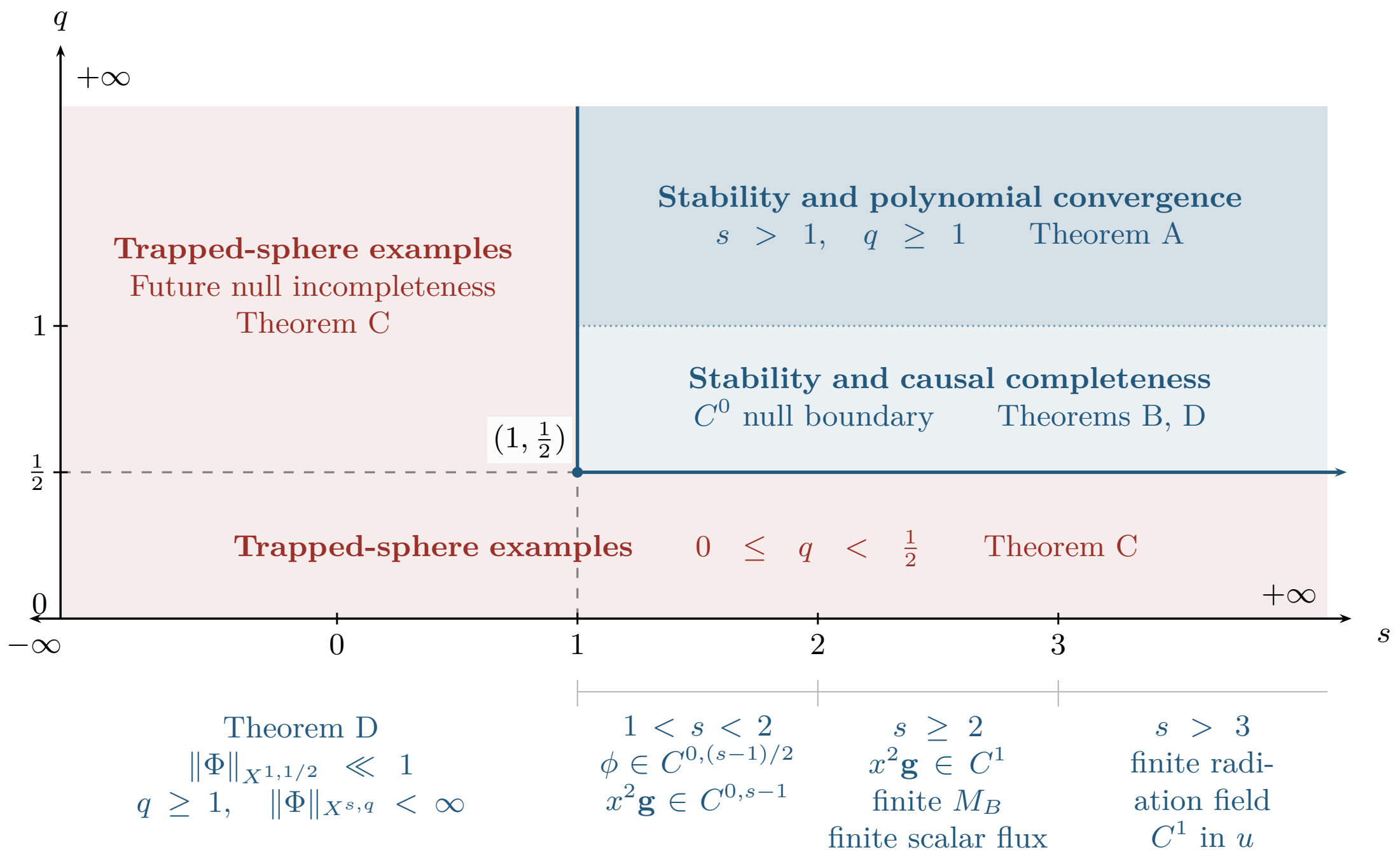

Figure~\ref{fig:parameter-regions} summarizes these thresholds and their geometric consequences.
Luk--Oh--Yang~\cite[Corollary~1.4 and Remark~5.1]{LOY} constructed future causally geodesically complete solutions with infinite Bondi mass on every cut of future null infinity and $\phi\to0$ along every outgoing cone by~\cite[Theorem~1.1, (1.9)]{LOY}. In the example of Theorem~\ref{thm:null-infinity}\textup{(iv)(a)}, $\phi$ has no limit along any outgoing cone.
\subsection{Relation to the vacuum stability problem}\label{sec:vacuum-problem}
To discuss the relation of our results to the Minkowski stability problem of Einstein vacuum equations, we first consider the Raychaudhuri equation on an outgoing cone:
\[
L(\operatorname{tr}\chi)+\frac12(\operatorname{tr}\chi)^2=-|\widehat\chi|^2-2|L\phi|^2.
\]
On $C_1$, a rough comparison between data in spherical symmetry and the shear tensor in vacuum is
\[
\Phi-\phi|_{C_1}=(r\pr_r\phi)|_{C_1}\quad\longleftrightarrow\quad r\widehat\chi.
\]
Linearizing the vacuum shear equation $(\nabla_L+2/r)\widehat\chi=-\alpha$ at Minkowski gives
\[
2\Phi'(v)=\left.\pr_r^2(r\phi)\right|_{C_1}=\left.r(\pr_r+2/r)\pr_r\phi\right|_{C_1}\quad\longleftrightarrow\quad -r\alpha.
\]
With $v=1+2r$ and $d\mu_{C_1}\asymp r^2drd\omega$, the last comparison tells us
\[
\|\Phi\|_{X^{s,1}}^2=\int_1^\infty v^s|\Phi'(v)|^2dv\quad\longleftrightarrow\quad\int_{C_1}(1+r)^s|\a|^2d\mu_{C_1}.
\]
At $s=1$ the scalar $L^2$ energy controls the quadratic radial sources. In vacuum, global stability of Minkowski at the corresponding $r$ weight remains open; see~\cite{ShenSurvey}.

For radial $f,h\in C_c^\infty(\mathbb R^3)$, extend $rf$ and $rh$ oddly to $\mathbb R$. The radial Fourier transform gives
\[
[\pr_r(rf)]_{\dot H^{1/2}(\mathbb R)}\asymp \|f\|_{\dot H^{3/2}(\mathbb R^3)},\qquad
[rh]_{\dot H^{1/2}(\mathbb R)}\asymp \|h\|_{\dot H^{1/2}(\mathbb R^3)}.
\]
Thus the critical regularity $q=1/2$ in Theorem~\ref{thm:half-stability} corresponds, at the level of derivative counting, to the scale-invariant Sobolev orders $\dot H^{3/2}\times\dot H^{1/2}$ for Cauchy data. This is closely related to Klainerman's strong stability conjecture for Minkowski spacetime~\cite[Lecture III, Section~3.1]{KlainermanWave}, which asks whether a scale-invariant smallness condition on general vacuum initial data $(\Sigma,g,k)$, involving locally $L^2$ control of $3/2$ derivatives of $g$ and $1/2$ derivatives of $k$, guarantees future timelike and null geodesic completeness of the maximal future development.

The bounded $L^2$ curvature theorem~\cite{KRS} establishes local vacuum evolution under $H^2\times H^1$ regularity, subject to suitable geometric assumptions. Its lifespan is controlled by the initial $L^2$ norms of curvature and $\nab k$, together with a lower bound on the volume radius. Local well-posedness at the critical Sobolev orders remains a separate question.

Related local and global questions were formulated by Luk--Moschidis~\cite[Problem~2.1]{LukMoschidis}. For $q\geq 1/2$, they ask whether vacuum Cauchy data arbitrarily small in $H^{q+1}(\mathbb{R}^3)\times H^q(\mathbb{R}^3)$, or suitable weighted or Besov variants, can lose control in finite time or lead to a future causally geodesically incomplete maximal development. Their formulation also encompasses arbitrarily high Sobolev regularity $q\gg1$, highlighting the question of global stability under unweighted Sobolev assumptions, independently of the optimal regularity threshold.

\subsection{Acknowledgements and AI statement}
{\it Acknowledgements:}
The authors thank Xuantao Chen, Sergiu Klainerman, and J\'er\'emie Szeftel for helpful discussions. J.W. is supported by ERC-2023 AdG 101141855 BlaHSt.

\medskip
{\it Disclosure of AI use:}
The main ideas of the proof, including the use of an auxiliary weight function and a scale-invariant energy--bulk estimate, are due to the authors. ChatGPT assisted with technical calculations, the presentation of the manuscript, and the continuation arguments in the appendix. The authors verified the arguments and calculations and take full responsibility for the results and the final manuscript.
\section{Preliminaries}\label{sec:pre}
We use the derivative
\[
\pr_r:=\la^{-1}\pr_v,
\]
to differentiate with respect to area radius along each $C_u$. We take $\pr_u$ at fixed $v$.
\subsection{Transport and radial identities}
The radial identities in a regular untrapped region use the coefficient
\begin{equation}\label{eq:G-definition}
 G:=\frac{\nu\mu}{(1-\mu)r}.
\end{equation}
\begin{prop}\label{lem:identities}
Every regular untrapped solution satisfies
\begin{align}
\pr_r m&=\frac12(1-\mu)r^2(\pr_r\phi)^2,\label{eq:mass-radial}\\
\pr_u\log\lambda&=G,\label{eq:lambda-u}\\
\pr_r\nu&=G,\label{eq:nu-radial}\\
\pr_u\bigl(\lambda^2\pr_r^2(r\phi)\bigr)&=\lambda^2\left[3G\pr_r\phi-\frac{\nu r}{1-\mu}(\pr_r\phi)^3\right].\label{eq:pruprrrphi}
\end{align}
\end{prop}
\begin{proof}
The mass equation in \eqref{eq:ssesf} gives \eqref{eq:mass-radial}. The radius equation gives
\[
 \pr_u\lambda=\pr_v\nu=\lambda G,
\]
which proves \eqref{eq:lambda-u} and \eqref{eq:nu-radial} and implies
\[
 [\pr_u,\pr_r]=-G\pr_r.
\]
Since $\mu=2m/r$, we also have
\[
 \pr_r\mu=(1-\mu)r(\pr_r\phi)^2-\frac{\mu}{r}.
\]
Differentiating \eqref{eq:G-definition} and using \eqref{eq:nu-radial}, we obtain
\[
 \pr_rG
 =\frac{G\mu}{(1-\mu)r}
  +\frac{\nu\pr_r\mu}{(1-\mu)^2r}
  -\frac{G}{r}
 =\frac{\nu}{1-\mu}(\pr_r\phi)^2-\frac{2G}{r}.
\]
Commuting $\pr_r$ with $\pr_u$ in $\pr_r\pr_u(r\phi)=G\phi$ gives
\[
 \pr_u\pr_r(r\phi)
 =G\phi-G\pr_r(r\phi)
 =-rG\pr_r\phi.
\]
Commute again and substitute $\pr_r^2(r\phi)=2\pr_r\phi+r\pr_r^2\phi$
\begin{align*}
 \pr_u\pr_r^2(r\phi)
 &=-G\pr_r^2(r\phi)-\pr_r(rG\pr_r\phi)\\
 &=-2G\pr_r^2(r\phi)+(G-r\pr_rG)\pr_r\phi\\
 &=-2G\pr_r^2(r\phi)+3G\pr_r\phi
   -\frac{\nu r}{1-\mu}(\pr_r\phi)^3.
\end{align*}
The last identity follows by multiplying by $\lambda^2$ and using $\pr_u(\lambda^2)=2G\lambda^2$.
\end{proof}
\begin{prop}\label{half:prop:radial-identities}
Every regular untrapped solution satisfies
\begin{align}
 \pr_u\pr_r(r\phi)&=-rG\pr_r\phi,
 \label{half:eq:h-transport}\\
 \pr_r(r^2G)
 &=\frac{\nu r^2}{1-\mu}(\pr_r\phi)^2,
 \label{half:eq:G-radial}\\
 \pr_r\log\frac{-\nu}{1-\mu}
 &=r(\pr_r\phi)^2.
 \label{half:eq:metric-log}
\end{align}
\begin{equation}\label{eq:null-constraints}
 \begin{aligned}
 \pr_u\nu-\nu\pr_u\log\Om^2&=-r|\pr_u\phi|^2,\\
 \pr_v\la-\la\pr_v\log\Om^2&=-r|\pr_v\phi|^2.
 \end{aligned}
\end{equation}
\end{prop}

\begin{proof}
The identity~\eqref{half:eq:h-transport}
was obtained in the proof of Proposition~\ref{lem:identities}.
The expression for $\pr_rG$ computed there gives
\[
\pr_r(r^2G)=2rG+r^2\pr_rG=\frac{\nu r^2}{1-\mu}(\pr_r\phi)^2,
\]
proving~\eqref{half:eq:G-radial}. Similarly, we have
\[
 \pr_r\log\frac{-\nu}{1-\mu}
 =\frac{G}{\nu}+\frac{\pr_r\mu}{1-\mu}
 =\frac{\mu}{(1-\mu)r}
 +r(\pr_r\phi)^2-\frac{\mu}{(1-\mu)r}
 =r(\pr_r\phi)^2.
\]
Finally,~\eqref{eq:geometry-definitions} and~\eqref{eq:ssesf} give
\begin{align*}
 \pr_u\nu-\nu\pr_u\log\Om^2
 &=-\nu G-\frac{\nu\pr_u\mu}{1-\mu} =-\frac{2\nu\pr_um}{(1-\mu)r}=-r|\pr_u\phi|^2,\\
 \pr_v\la-\la\pr_v\log\Om^2
 &=-\frac{\la^2G}{\nu}-\frac{\la\pr_v\mu}{1-\mu} =-\frac{2\la\pr_vm}{(1-\mu)r}=-r|\pr_v\phi|^2.
\end{align*}
\end{proof}
\subsection{Scaling and criticality}
\label{sec:scaling-criticality}
For $A>0$, put $\tilde{u}=1+(u-1)/A$, $\tilde{v}=1+(v-1)/A$, and consider the rescaling
\begin{equation}\label{eq:dilation}
\begin{aligned}
r_A(u,v)&=A r(\tilde u,\tilde v),&m_A(u,v)&=A m(\tilde u,\tilde v),\\
(\phi_A,\la_A,\nu_A,\mu_A,\Omega_A)(u,v)&=(\phi,\la,\nu,\mu,\Omega)(\tilde u,\tilde v),&G_A(u,v)&=A^{-1}G(\tilde u,\tilde v).
\end{aligned}
\end{equation}
It's straightforward to check that~\eqref{eq:dilation} preserves~\eqref{eq:ssesf}, the gauge and axis conditions, and the initial geometric bounds. The rescaled initial data is
\begin{equation}\label{eq:dilated-data}
\Phi_A(v)=\Phi\left(1+\frac{v-1}{A}\right).
\end{equation}
\begin{lem}\label{lem:log-norm-equivalence}\label{lem:subcritical-scaling}
Let $q>0$, $0\le c_0<c_1$, and $A\ge1$, with $c_0>0$ and $s>1/2$ when $A>1$ (and $s\in\mathbb R$ when $A=1$). Put $\tau=\log v$ and $f_A(\tau)=\Phi_A(e^\tau)$. If $\Phi_A$ is smooth and supported in $[1+c_0 A,1+c_1 A]$, then
\begin{equation}\label{eq:log-norm-equivalence}
\|\Phi_A\|_{X^{s,q}}^2\asymp_{s,q,c_0,c_1} A^{s-1}\|f_A\|_{\dot{H}^q(\mathbb R_+)}^2,
\end{equation}
where $\mathbb R_+=(0,\infty)$ and the fractional seminorm is taken over $\mathbb R_+^2$.
\end{lem}
\begin{proof}
Since $v\pr_v=\pr_\tau$ and $dv=e^\tau d\tau$, $\operatorname{supp} f_A$ lies in $I_A:=[\log(1+c_0 A),\log(1+c_1 A)]$, with
\begin{equation}\label{eq:IA-bounds}
|I_A|=\log\frac{1+c_1 A}{1+c_0 A}\les_{c_0,c_1}1,\qquad e^\tau\asymp_{c_0,c_1} A\quad (\tau\in I_A).
\end{equation}
For every integer $m\ge1$, \eqref{eq:IA-bounds} gives
\begin{equation}\label{eq:integer-tau-equiv}
\int_1^\infty v^{s-1}|(v\pr_v)^m\Phi_A(v)|^2\frac{dv}{v}
=\int_{I_A} e^{(s-1)\tau}|\pr_\tau^m f_A(\tau)|^2\,d\tau
\asymp_{s,c_0,c_1} A^{s-1}\int_{I_A}|\pr_\tau^m f_A(\tau)|^2\,d\tau.
\end{equation}
For $q=k+\beta$ ($k\in\mathbb N_0$, $0<\beta<1$), write
\[
J_{s,q}(\Phi_A):=\int_1^\infty\int_1^\infty
\frac{\min\{v,v'\}^{s-1}(vv')^{\beta-1/2}}{|v-v'|^{1+2\beta}}
\left|(v\pr_v)^k\Phi_A(v)-(v'\pr_{v'})^k\Phi_A(v')\right|^2dv\,dv'.
\]
Changing variables $v=e^\tau$ and $v'=e^{\tau'}$, and using $\pr_{\tau'}^k f_A(\tau')=0$ for $\tau'\in\mathbb R_+\setminus I_A$, gives
\begin{equation}\label{eq:J-split}
J_{s,q}(\Phi_A)=J_{s,q}^{\mathrm{diag}}(\Phi_A)+J_{s,q}^{\mathrm{tail}}(\Phi_A),
\end{equation}
where
\begin{align*}
J_{s,q}^{\mathrm{diag}}(\Phi_A)
&:=\iint_{I_A\times I_A}
 e^{(s-1)\min\{\tau,\tau'\}}
 \frac{e^{-(\beta+1/2)|\tau-\tau'|}}{\big(1-e^{-|\tau-\tau'|}\big)^{1+2\beta}}
 \left|\pr_\tau^k f_A(\tau)-\pr_{\tau'}^k f_A(\tau')\right|^2d\tau\,d\tau',\\
J_{s,q}^{\mathrm{tail}}(\Phi_A)
&:=2\int_{I_A}\left(\int_{\mathbb R_+\setminus I_A}
 e^{(s-1)\min\{\tau,\tau'\}}
 \frac{e^{-(\beta+1/2)|\tau-\tau'|}}{\big(1-e^{-|\tau-\tau'|}\big)^{1+2\beta}}
 \,d\tau'\right)\left|\pr_\tau^k f_A(\tau)\right|^2d\tau.
\end{align*}
By estimating the weights in the integrals, one can easily show that
\begin{align}
J_{s,q}^{\mathrm{diag}}(\Phi_A)
&\asymp_{s,c_0,c_1} A^{s-1}\iint_{I_A\times I_A}
\frac{\left|\pr_\tau^k f_A(\tau)-\pr_{\tau'}^k f_A(\tau')\right|^2}{|\tau-\tau'|^{1+2\beta}}
\,d\tau\,d\tau',\label{eq:J-diag-est}\\
J_{s,q}^{\mathrm{tail}}(\Phi_A)
&\asymp_{s,c_0,c_1} A^{s-1}\int_{I_A}\left|\pr_\tau^k f_A(\tau)\right|^2
\left(1+\int_{\substack{\tau'\in\mathbb R_+\setminus I_A\\|\tau-\tau'|\le1}}
\frac{d\tau'}{|\tau-\tau'|^{1+2\beta}}\right)d\tau.
\label{eq:J-tail-est}
\end{align}
Now we show both bounds in~\eqref{eq:log-norm-equivalence}. Write $b_A=\sup I_A$. Since $\pr_\tau^m f_A(b_A)=0$, by Poincar\'e we have
\[
\|\pr_\tau^m f_A\|_{L^2(I_A)}^2
\le\frac{|I_A|^2}{2}\|\pr_\tau^{m+1} f_A\|_{L^2(I_A)}^2,\qquad m\ge0.
\]
Thus, for integer $q\ge1$, \eqref{eq:integer-tau-equiv} gives
\begin{align*}
A^{s-1}\|\pr_\tau^q f_A\|_{L^2(I_A)}^2
&\les_{s,c_0,c_1}\|\Phi_A\|_{X^{s,q}}^2\\
&\les_{s,c_0,c_1}A^{s-1}
   \sum_{m=1}^q\|\pr_\tau^m f_A\|_{L^2(I_A)}^2\\
&\le A^{s-1}\|\pr_\tau^q f_A\|_{L^2(I_A)}^2
   \sum_{m=1}^q\left(\frac{|I_A|^2}{2}\right)^{q-m}.
\end{align*}
For $q=k+\beta$, put $g=\pr_\tau^k f_A$. The unweighted tail with $|\tau-\tau'|>1$ is comparable to $\|g\|_{L^2(I_A)}^2$ for fixed $\beta$. Thus~\eqref{eq:J-diag-est} and~\eqref{eq:J-tail-est} imply
\[
A^{s-1}\|f_A\|_{\dot H^q(\mathbb R_+)}^2
\les_{s,q,c_0,c_1}
J_{s,q}^{\mathrm{diag}}(\Phi_A)+J_{s,q}^{\mathrm{tail}}(\Phi_A)
\le \frac{1}{1-\beta}\|\Phi_A\|_{X^{s,q}}^2.
\]
Conversely, combining~\eqref{eq:integer-tau-equiv} with~\eqref{eq:J-split}--\eqref{eq:J-tail-est} gives
\begin{align*}
\|\Phi_A\|_{X^{s,q}}^2
&\les_{s,q,c_0,c_1} A^{s-1}\left(
\sum_{m=1}^k\int_{I_A}|\pr_\tau^m f_A(\tau)|^2\,d\tau
+\iint_{\mathbb R_+^2}
\frac{|g(\tau)-g(\tau')|^2}{|\tau-\tau'|^{1+2\beta}}\,
d\tau\,d\tau'\right)\\
&\les_{s,q,c_0,c_1} A^{s-1}\|f_A\|_{\dot H^q(\mathbb R_+)}^2.
\end{align*}
For the last inequality, since $g=0$ on $[b_A+1,b_A+2]$,
\begin{align*}
[g]_{\dot H^\beta(\mathbb R_+)}^2\ge\int_{I_A}\int_{b_A+1}^{b_A+2}
    \frac{|g(\tau)|^2}{|\tau-\tau'|^{1+2\beta}}\,d\tau'\,d\tau\ge (|I_A|+2)^{-1-2\beta}\|g\|_{L^2(I_A)}^2.
\end{align*}
Therefore
\[
\sum_{m=1}^k\|\pr_\tau^m f_A\|_{L^2(I_A)}^2
\le \|g\|_{L^2(I_A)}^2
   \sum_{m=1}^k\left(\frac{|I_A|^2}{2}\right)^{k-m}
\les_{q,c_0,c_1}[g]_{\dot H^\beta(\mathbb R_+)}^2.
\]
This proves~\eqref{eq:log-norm-equivalence}.
\end{proof}
\begin{prop}\label{prop:scaling}
Let $0\ne\Phi\in C_c^\infty((2,3))$ and define $\Phi_A$ by~\eqref{eq:dilated-data} for $A>0$.
\begin{enumerate}
\item[\textup{(i)}] Dilation $A\ge1$: for $1/2<s<1$ and an integer $M\ge1$,
\begin{equation}\label{eq:subcritical-scaling}
\sup_{0\le q\le M}\|\Phi_A\|_{X^{s,q}}^2\les_{s,M}\|\Phi\|_{C^M}^2 A^{s-1}.
\end{equation}
\item[\textup{(ii)}] Concentration $0<A\le1$: for $s\in\mathbb R$ and $q>0$,
\begin{equation}\label{eq:X-concentration-scaling}
\|\Phi_A\|_{X^{s,q}}^2\asymp_{\Phi,s,q}A^{1-2q}.
\end{equation}
\item[\textup{(iii)}] Criticality at $(s,q)=(1,1/2)$: for all $A>0$,
\begin{equation}\label{eq:X-critical-dilation}
 \|\Phi_A\|_{X^{1,1/2}}=\|\Phi\|_{X^{1,1/2}}.
\end{equation}
\end{enumerate}
\end{prop}
\begin{proof}
\noindent Case (i): For $A\ge1$, $\operatorname{supp}\Phi_A\subset[1+A,1+2A]$, and $f_A(\tau):=\Phi_A(e^\tau)=\Phi(1+(e^\tau-1)/A)$ satisfies
\[
f_A(\log A+\sigma)=\Phi\big(1+e^\sigma-A^{-1}\big),
\]
which is supported in $[0,\log 3]$ and satisfies $\|f_A(\log A+\cdot)\|_{C^M}\les_M \|\Phi\|_{C^M}$ uniformly for $A\ge1$. By~\eqref{eq:integer-tau-equiv} and~\eqref{eq:J-diag-est}--\eqref{eq:J-tail-est}, it's straightforward to show that
\begin{align*}
\sup_{0\le q\le M}\|\Phi_A\|_{X^{s,q}}^2
&\les_{s,M} A^{s-1}\|f_A\|_{H^M(\mathbb R_+)}^2
 \les_{s,M}\|\Phi\|_{C^M}^2 A^{s-1}.
\end{align*}
\noindent Case (ii): For $0<A\le1$, $\operatorname{supp}\Phi_A\subset[1+A,1+2A]\subset[1,3]$. Applying Lemma~\ref{lem:log-norm-equivalence} at scale $1$ with $c_0=0$ and $c_1=2$ gives $\|\Phi_A\|_{X^{s,q}}^2\asymp_{s,q}\|f_A\|_{\dot{H}^q(\mathbb R_+)}^2$. Setting $\tau=Ay$, we see that $f_A(Ay)=\Phi(1+(e^{Ay}-1)/A)$ is supported in $[\log 2,2]$ and converges in $C_c^\infty(\mathbb R)$ as $A\downarrow0$ to $y\mapsto\Phi(1+y)$. Scaling $\dot{H}^q(\mathbb R_+)$ gives
\[
\|f_A\|_{\dot{H}^q(\mathbb R_+)}^2
=A^{1-2q}\|f_A(A \cdot)\|_{\dot{H}^q(\mathbb R_+)}^2
\asymp_{\Phi,q}A^{1-2q},
\]
which proves~\eqref{eq:X-concentration-scaling}.

\noindent Case (iii): Changing variables $v=1+Ax$ and $v'=1+Ax'$ in~\eqref{half:eq:data-norm} gives directly~\eqref{eq:X-critical-dilation}, which shows the criticality at $(s,q)=(1,1/2)$.
\end{proof}
\begin{rk}
For $A\ge1$ and $q>0$, one can also show $\|\Phi_A\|_{X^{s,q}}^2\asymp_{\Phi,s,q} A^{s-1}$ for $s>1/2$ and an upper bound for $s\le1/2$. Since we will not use these bounds later, we omit their proofs.
\end{rk}
\section{Stability for one derivative at critical decay threshold}\label{sec:stab1}
Fix $s\ge1$ and initial data satisfying~\eqref{eq:data} and
\[
\int_1^\infty v^s|\Phi'(v)|^2dv\le\eps_0^2.
\]
Set $\epsilon=\epsilon_0^{2/3}$. Implicit constants may depend on $s$ and are uniform in $\epsilon$, $\epsilon_0$, the data, and the size of the triangle $\mathcal T_V=\{(u,v):1\le u\le v\le V\}$.
\subsection{Initial bounds and bootstrap assumptions}
\begin{lem}\label{envelope}
Suppose that $\Phi\in C^1([1,\infty))$ is nonconstant and
\[
\int_1^\infty z|\Phi'(z)|^2dz\le\epsilon_0^2,\qquad \eps_0>0.
\]
Then there exists a continuous function $L:[1,\infty)\to[1,\infty)$, depending only on $\Phi$ and $\epsilon_0$, such that $L(v)\to\infty$ as $v\to\infty$ and
\begin{align}
\int_1^v z^{3/2}|\Phi'(z)|^2dz&\le\frac{\epsilon_0^2v^{1/2}}{L(v)^2},\qquad v\ge1,\label{L-ID-relation}\\
\frac{L(x)}{L(y)}&\le\max\left\{\left(\frac xy\right)^{1/4},\left(\frac yx\right)^{1/4}\right\},\qquad x,y\ge1,\label{L-comparison}\\
\int_1^\infty\frac{dv}{vL(v)^2}&\les 1.\label{borderline-property}
\end{align}
\end{lem}
\begin{proof}
We construct an envelope of the initial derivative energy by setting
\begin{equation}\label{eq:one-derivative-L}
 L(v)^{-2}
 :=\epsilon_0^{-2}\int_1^\infty
 e^{-|\log(v/z)|/2}|z\Phi'(z)|^2\frac{dz}{z}.
\end{equation}
Since $\Phi$ is nonconstant and the kernel is positive and bounded by $1$,
\[
 0<L(v)^{-2}
 \le\epsilon_0^{-2}\int_1^\infty z|\Phi'(z)|^2dz
 \le1.
\]
Dominated convergence gives continuity and, since the kernel tends to zero as $v\to\infty$ for each fixed $z$, also gives $L(v)^{-2}\to0$. For $1\le z\le v$, the kernel equals $(z/v)^{1/2}$. Thus,
\[
 L(v)^{-2}\geq \epsilon_0^{-2}\int_1^v
 e^{-\log(v/z)/2}|z\Phi'(z)|^2\frac{dz}{z}\geq\epsilon_0^{-2}v^{-1/2}\int_1^v z^{3/2}|\Phi'(z)|^2dz,
\]
which proves \eqref{L-ID-relation}. By the triangle inequality,
\[
 e^{-|\log(x/z)|/2}
 \ge e^{-|\log(x/y)|/2}e^{-|\log(y/z)|/2}.
\]
Substituting this into \eqref{eq:one-derivative-L} yields
\[
 L(x)^{-2}
 \ge e^{-|\log(x/y)|/2}L(y)^{-2}.
\]
Consequently,
\[
 \frac{L(x)}{L(y)}
 \le e^{|\log(x/y)|/4}=\max\left\{\left(\frac xy\right)^{1/4},\left(\frac yx\right)^{1/4}\right\},
\]
proving \eqref{L-comparison}. Finally, for fixed $z\ge1$, direct integration gives
\begin{align*}
\int_1^\infty e^{-|\log(v/z)|/2}\frac{dv}{v}=\int_1^z\left(\frac vz\right)^{1/2}\frac{dv}{v}+\int_z^\infty\left(\frac zv\right)^{1/2}\frac{dv}{v}=4-2z^{-1/2}\le4.
\end{align*}
Tonelli's theorem therefore implies
\[
\int_1^\infty\frac{dv}{vL(v)^2}=\epsilon_0^{-2}\int_1^\infty|z\Phi'(z)|^2(4-2z^{-1/2})\frac{dz}{z}\le4,
\]
which proves \eqref{borderline-property}.
\end{proof}
\begin{rk}\label{rk:L-motivation}
The construction~\eqref{eq:one-derivative-L} is a logarithmic-scale analogue of Tao's frequency-envelope method~\cite[Section~2]{TaoEnvelopes}: exponential averaging preserves integrability while enforcing the scale comparison~\eqref{L-comparison}. The condition~\eqref{borderline-property} is also motivated by the borderline weights in the initial-data construction of \cite[Proposition~4.2 and Remark~4.3]{ShenWanIDS}. For a related forward construction of borderline vacuum initial data, see \cite[Section~1.1]{ChenKlainermanIDS}. Here $L$ is constructed from the given characteristic data.
\end{rk}
For $s=1$, let $L$ be constructed in \eqref{eq:one-derivative-L}. For $s>1$, we define
\begin{equation}\label{choiceLs>1}
L(v):=v^{\min\left\{\frac{s-1}{2},\frac14\right\}}.
\end{equation}
Both choices satisfy \eqref{L-ID-relation}--\eqref{borderline-property}, with $L\geq 1$ and $\lim_{v\to\infty}L(v)=\infty$.
\begin{lem}
Under the initial data assumptions, for sufficiently small $\epsilon_0$, we have on $C_1$
\[
\la=\frac{1}{2},\qquad\left|\nu+\frac12\right|+\mu\les\epsilon_0^2,\qquad|\pr_r\phi|\les\frac{\epsilon_0}{\sqrt{rv}\,L(v)}.
\]
\end{lem}
\begin{proof}
All quantities below are restricted to $C_1$, where $\la=1/2$ and $r=(v-1)/2$. The initial normalization and integration by parts give
\begin{align*}
 |\pr_r\phi|
 &=\frac2{(v-1)^2}
 \left|(v-1)\Phi(v)-\int_1^v\Phi(z)dz\right|\\
 &=\frac2{(v-1)^2}
 \left|\int_1^v(z-1)\Phi'(z)dz\right|\\
 &\le\frac2{(v-1)^2}
 \left(\int_1^v\frac{(z-1)^2}{z^{3/2}}dz\right)^{1/2}
 \left(\int_1^v z^{3/2}|\Phi'(z)|^2dz\right)^{1/2}\\
 &\les\frac{\epsilon_0v^{1/4}}{(v-1)^2L(v)}
 \left(\frac{v-1}{v}\right)^{3/4}
 \left(\int_1^v(z-1)^{1/2}dz\right)^{1/2}\\
 &\les\frac{\epsilon_0}{\sqrt{rv}\,L(v)}.
\end{align*}
Here we used Cauchy--Schwarz and \eqref{L-ID-relation}, then $(z-1)/z\le(v-1)/v$ for $1\le z\le v$, and finally $r=(v-1)/2$. Thus \eqref{borderline-property} implies
\[
 \int_0^\infty r|\pr_r\phi|^2dr\les\epsilon_0^2.
\]
The mass constraint \eqref{eq:mass-radial} implies $\pr_r m+r(\pr_r\phi)^2m=\frac12r^2(\pr_r\phi)^2$, with $m(0)=0$, yields
\[
 0\le m(r)\le\frac12\int_0^r {r'}^2|\pr_r\phi(r')|^2dr'.
\]
For the pointwise bound, use $\mu=2m/r$. For the integral of $\mu/r$, change the order of integration by Tonelli's theorem to get
\[
\sup_{r>0}\mu(r)+\int_0^\infty\frac{\mu(r)}rdr\les\int_0^\infty r|\pr_r\phi|^2dr\les\epsilon_0^2.
\]
For small $\epsilon_0$, $1-\mu\ge1/2$. Integrating~\eqref{eq:nu-radial} from the axis, where $\nu=-1/2$, gives
\[
0\le\log(-2\nu(r))=\int_0^r\frac{\mu(r')}{(1-\mu(r'))r'}dr'\les\epsilon_0^2.
\]
Exponentiating gives $|\nu+1/2|\les\epsilon_0^2$.
\end{proof}
\noindent\textbf{Bootstrap assumptions.}
On a finite characteristic triangle, assume that the regular untrapped solution satisfies
\begin{equation}\label{q=1B1}
 \left|\lambda-\frac12\right|+\left|\nu+\frac12\right|+\mu\le\epsilon,
 \qquad
 |\pr_r\phi|\le\frac{\epsilon}{\sqrt{rv}\,L(v)}. \tag{B.A.-Thm A}
\end{equation}
The bounds in~\eqref{q=1B1} and integration of $\la$ from the axis give
\[
\la\asymp1,\qquad -\nu\asymp1,\qquad 1-\mu\asymp1,\qquad r\asymp v-u.
\]
\subsection{Bootstrap estimates}
\begin{lem}\label{lem:radial-integral}
Under the bootstrap assumption \eqref{q=1B1}, for every $0<a<2$
\[
\int_0^r\frac{r'}{(v')^a}dr'\les\frac{r^2}{v^a},
\]
with $r:=r(u,v)$ and $r':=r(u,v')$ along the fixed cone $C_u$.
\end{lem}
\begin{proof}
At fixed $u$, the bound $r\asymp v-u$ gives $v'\asymp u+r'$ and $v\asymp u+r$. The ratio $r'/(u+r')$ increases with $r'$, so
\begin{align*}
 \int_0^r\frac{r'}{(v')^a}dr'\les\int_0^r\left(\frac{r'}{u+r'}\right)^a(r')^{1-a}dr'\le\left(\frac r{u+r}\right)^a\int_0^r(r')^{1-a}dr'\les\frac{r^2}{(u+r)^a}\les\frac{r^2}{v^a}
\end{align*}
as stated.
\end{proof}
\begin{prop}\label{prop:mass-mu-G}
Under the above assumptions, for sufficiently small $\epsilon_0$, we have
\[
 0\le m\les\frac{\epsilon^2r^2}{vL(v)^2},
 \qquad 0\le\mu\les\frac{\epsilon^2r}{vL(v)^2},
 \qquad 0\le-G\les\frac{\epsilon^2}{vL(v)^2}.
\]
\end{prop}
\begin{proof}
Integrate~\eqref{eq:mass-radial} from the axis and insert the scalar bound in~\eqref{q=1B1}. The comparison~\eqref{L-comparison} then puts the integral in the form of Lemma~\ref{lem:radial-integral} with $a=3/2$.
\[
 0\le m(u,r)
 \les\int_0^r(r')^2|\pr_r\phi(u,r')|^2dr'
 \les\epsilon^2\int_0^r\frac{r'}{v'L(v')^2}dr'
 \les\frac{\epsilon^2v^{1/2}}{L(v)^2}
       \int_0^r\frac{r'}{(v')^{3/2}}dr'
 \les\frac{\epsilon^2r^2}{vL(v)^2}.
\]
The estimates for $\mu$ and $G$ now follow from $\mu=2m/r$ and~\eqref{eq:G-definition}.
\end{proof}
\begin{prop}\label{prop:lambda-nu}
Under the above assumptions, for sufficiently small $\epsilon_0$, we have
\[
 0\le\frac12-\la\les\frac{\epsilon_0}{L(v)^2},\qquad
 \left|\nu+\frac12\right|+|\Om^2-1|
 \les\epsilon_0\left(
 \frac1{L(u)^2}+\int_u^v\frac{dz}{zL(z)^2}\right).
\]
\end{prop}
\begin{proof}
Using the bound for $G$ in Proposition~\ref{prop:mass-mu-G}, integrating $\pr_u\log\la=G$ from \eqref{eq:lambda-u}, with $\la(1,v)=1/2$, gives
\[
 0\le\frac12-\la(u,v)
 \le-\frac12\log(2\la(u,v))
 =-\frac12\int_1^uG(u',v)du'
 \les\frac{\epsilon^2(u-1)}{vL(v)^2}
 \le\frac{\epsilon^2}{L(v)^2}.
\]
Integrating $\pr_r\nu=G$ from \eqref{eq:nu-radial} and using $\nu(u,0)=-\la(u,0)$ at the regular axis, we obtain
\[
 \left|\nu(u,r)+\frac12\right|
 =\left|\frac12-\la(u,0)+\int_0^rG(u,r')dr'\right|
 \les\epsilon^2\left(
 \frac1{L(u)^2}+\int_u^v\frac{dz}{zL(z)^2}\right).
\]
Moreover, for $u\le z\le v$, \eqref{L-comparison} gives
\[
 L(z)^{-2}\ge(z/v)^{1/2}L(v)^{-2}.
\]
Consequently,
\[
L(u)^{-2}+\int_u^v\frac{dz}{zL(z)^2}\geq\left(2-\sqrt{\frac{u}{v}}\right)L(v)^{-2}\geq L(v)^{-2}.
\]
Using $\Om^2=-4\la\nu/(1-\mu)$, the metric bootstrap, and Proposition~\ref{prop:mass-mu-G}, we therefore find
\[
|\Om^2-1|\les\left|\la-\frac12\right|+\left|\nu+\frac12\right|+\mu\les\epsilon^2\left(\frac1{L(u)^2}+\int_u^v\frac{dz}{zL(z)^2}\right).
\]
Since $\epsilon^2=\epsilon_0^{4/3}\le\epsilon_0$, the stated estimates follow.
\end{proof}
\begin{lem}\label{lem:q1-null-variation}
Let $v\ge1$, $u_1<u$, and $\epsilon,L(v)>0$. Suppose that
\[
 \lambda,-\nu,1-\mu\asymp1,\qquad r\les v,\qquad
 |\pr_r\phi|\les\frac{\epsilon}{\sqrt{rv}\,L(v)},\qquad
 |G|\les\frac{\epsilon^2}{vL(v)^2}
\]
on the incoming null segment $[u_1,u]\times\{v\}$. Then
\begin{equation}\label{eq:q1-null-variation}
 \int_{u_1}^u\left|\pr_u\bigl(\lambda^2\pr_r^2(r\phi)\bigr)(u',v)\right|du'
 \les\frac{\epsilon^3}{vL(v)^3}.
\end{equation}
\end{lem}
\begin{proof}
By~\eqref{eq:pruprrrphi},
\[
 \pr_u\bigl(\lambda^2\pr_r^2(r\phi)\bigr)
 =\lambda^2\left[3G\pr_r\phi-\frac{\nu r}{1-\mu}(\pr_r\phi)^3\right].
\]
Since $-\pr_ur=-\nu\asymp1$,
\begin{align*}
 \int_{u_1}^u\left|\pr_u\bigl(\lambda^2\pr_r^2(r\phi)\bigr)(u',v)\right|du'
 &\les\int_{u_1}^u\bigl(|G\pr_r\phi|+r|\pr_r\phi|^3\bigr)(u',v)du'\\
 &\les\frac{\epsilon^3}{v^{3/2}L(v)^3}
       \int_{r(u,v)}^{r(u_1,v)}\frac{dr'}{\sqrt{r'}}
 \les\frac{\epsilon^3}{vL(v)^3}.
\end{align*}
\end{proof}
\begin{prop}
\label{prop:radial-scalar}
Under the above assumptions, for sufficiently small $\epsilon_0$, we have
\[
|\pr_r\phi(u,v)|\les\frac{\epsilon_0}{\sqrt{rv}\,L(v)}.
\]
\end{prop}
\begin{proof}
By Lemma~\ref{lem:q1-null-variation} and $\pr_r^2(r\phi)(1,v)=2\Phi'(v)$,
\begin{equation}\label{prr2rphi}
 |\pr_r^2(r\phi)(u,v)|
 \les|\Phi'(v)|+\frac{\epsilon^3}{vL(v)^3}.
\end{equation}
By $\pr_r(r^2\pr_r\phi)=r\pr_r^2(r\phi)$, integration in $r$ along $C_u$ gives
\begin{align}\label{prrphi-est}
|\pr_r\phi(u,v)|\le\frac1{r^2}\int_0^rr'|\pr_r^2(r\phi)(u,r')|dr' \les\frac1{r^2}\int_0^r r'|\Phi'(v')|dr'+\frac{\epsilon^3}{r^2}\int_0^r\frac{r'}{v'L(v')^3}dr'.
\end{align}
For the initial data term, \eqref{L-ID-relation} and $dr'=\la(u,v')dv'$ imply
\[
\int_0^r(v')^{3/2}|\Phi'(v')|^2dr'\les\int_u^v (v')^{3/2}|\Phi'(v')|^2dv'\le\frac{\epsilon_0^2v^{1/2}}{L(v)^2}.
\]
Lemma~\ref{lem:radial-integral} with $a=3/2$ gives
\[
\int_0^r\frac{(r')^2}{(v')^{3/2}}dr'\le r\int_0^r\frac{r'}{(v')^{3/2}}dr'\les\frac{r^3}{v^{3/2}}.
\]
Cauchy--Schwarz therefore yields
\begin{align}\label{prrphi-ID}
\int_0^r r'|\Phi'(v')|dr'\le\left(\int_0^r(v')^{3/2}|\Phi'(v')|^2dr'\right)^{1/2}\left(\int_0^r\frac{(r')^2}{(v')^{3/2}}dr'\right)^{1/2} \les\frac{\epsilon_0r^{3/2}}{\sqrt v\,L(v)}.
\end{align}
For the nonlinear term, \eqref{L-comparison} and Lemma~\ref{lem:radial-integral} with $a=7/4$ give
\begin{equation}\label{prrphi-NL}
\int_0^r\frac{r'}{v'L(v')^3}dr'\le\frac{v^{3/4}}{L(v)^3}\int_0^r\frac{r'}{(v')^{7/4}}dr'\les\frac{r^2}{vL(v)^3}.
\end{equation}
Substituting estimates \eqref{prrphi-ID} and \eqref{prrphi-NL} into \eqref{prrphi-est}, using $r\les v$ and $L(v)\ge1$, we infer
\[
|\pr_r\phi(u,v)|\les\frac{\eps_0}{\sqrt{rv}\,L(v)}+\frac{\eps^3}{vL(v)^3}
\les\frac{\eps_0+\eps^3}{\sqrt{rv}\,L(v)}\les\frac{\eps_0}{\sqrt{rv}\,L(v)},
\]
which concludes the proof.
\end{proof}
\subsection{Proof of Theorem~\ref{thm:s=1}}\label{sec:proofA}
\noindent\textbf{Step 1. Bootstrap closure and global existence.}
In Proposition~\ref{prop:lambda-nu}, use $L\ge1$ and bound the integral by~\eqref{borderline-property}. Together with Propositions~\ref{prop:mass-mu-G} and~\ref{prop:radial-scalar}, this gives
\[
 \left|\la-\frac12\right|+\left|\nu+\frac12\right|
 +|\Omega^2-1|+\mu\les\epsilon_0\leq\frac{\eps}{2},\qquad
 \sqrt{rv}\,L(v)|\pr_r\phi|\les\epsilon_0\leq\frac{\eps}{2},
\]
Both bounds improve~\eqref{q=1B1} uniformly in the triangle size, closing the bootstrap by continuity. Since $\la\sqrt{r/v}/L(v)\les1$, the scalar estimate also gives $|r\pr_v\phi|\les\epsilon_0$, completing~\eqref{eq:geometry-main}.

The local well-posedness theory of~\cite[Section~2.1]{LOY} and the improved bootstrap bounds yield the unique global smooth solution. See Proposition~\ref{app:prop:continuation}.

\noindent\textbf{Step 2. Future causal geodesic completeness.}\label{step:causal-completeness}
The scalar equation in~\eqref{eq:ssesf} and~\eqref{eq:nu-radial} give
\begin{equation}\label{eq:incoming-scalar}
\pr_r\!\left(\frac r\nu\pr_u\phi\right)+\frac G\nu\frac r\nu\pr_u\phi=-\pr_r\phi.
\end{equation}
The integrating factor is $-\nu$, since $\pr_r\log(-\nu)=G/\nu$. The boundary term vanishes at the regular axis, giving
\begin{equation}\label{eq:incoming-scalar-integral}
 r\pr_u\phi(u,v)=-\int_0^r\nu(u,r')\pr_r\phi(u,r')dr'.
\end{equation}
Since $-\nu\asymp1$, Proposition~\ref{prop:radial-scalar} and~\eqref{L-comparison} give
\[
 |\pr_u\phi(u,v)|
 \les\frac1r\int_0^r|\pr_r\phi(u,r')|dr'
 \les\frac{\epsilon_0}{\sqrt{ru}\,L(u)}.
\]
For the geodesic estimate we use the bound
$$
r|\pr_u\phi|^2\les\epsilon_0^2/(uL(u)^2).
$$

For an affinely parametrized future causal geodesic $\tau\mapsto(u(\tau),v(\tau),\omega(\tau))$, write dots for $d/d\tau$ and set $\mathcal E=\la\dot v-\nu\dot u$. The geodesic equations give
\begin{align*}
 \ddot u+\pr_u\log\Om^2\,\dot u^2&=-\frac{2r\la}{\Om^2}|\dot\omega|_{\mathbb S^2}^2,\\
 \ddot v+\pr_v\log\Om^2\,\dot v^2&=-\frac{2r\nu}{\Om^2}|\dot\omega|_{\mathbb S^2}^2.
\end{align*}
Since $\pr_u\la=\pr_v\nu$, the null constraints~\eqref{eq:null-constraints} yield
\begin{align*}
 \dot{\mathcal E}
 &=(\pr_v\la)\dot v^2-(\pr_u\nu)\dot u^2+\la\ddot v-\nu\ddot u\\
 &=(\pr_v\la-\la\pr_v\log\Om^2)\dot v^2
 -(\pr_u\nu-\nu\pr_u\log\Om^2)\dot u^2\\
 &=r|\pr_u\phi|^2\dot u^2-r|\pr_v\phi|^2\dot v^2\les\frac{\epsilon_0^2}{uL(u)^2}\dot u\mathcal E.
\end{align*}
Hence, by~\eqref{borderline-property},
\[
 \mathcal E(\tau)\le\mathcal E(0)\exp\!\left(C\epsilon_0^2\int_{u(0)}^{u(\tau)}\frac{du}{uL(u)^2}\right)\les\mathcal E(0).
\]
Causality gives
\[
r^2|\dot\omega|_{\mathbb S^2}^2\le\Om^2\dot u\dot v,\qquad\dot u+\dot v+|\dot r|+r|\dot\omega|_{\mathbb S^2}\les\mathcal E.
\]
Thus the spacetime is future causally geodesically complete.

\noindent\textbf{Step 3. Derivative energy and convergence.}
By \eqref{prr2rphi}, \eqref{borderline-property}, \eqref{eq:small}, and $r\les v$,
\[
 \int_u^\infty r|\pr_r^2(r\phi)(u,v)|^2\la\,dv
 \les\int_u^\infty v|\Phi'(v)|^2dv+\eps_0^4\int_u^\infty\frac{dv}{vL(v)^6}\les\eps_0^2.
\]
Both tails tend to zero as $u\to\infty$, proving \eqref{eq:one-derivative-energy} and \eqref{eq:energy-limit}. The geometric estimates give
\[
\sup_{v>u}\left(\left|\la-\frac12\right|+\left|\nu+\frac12\right|+|\Omega^2-1|+\mu\right)\les\epsilon_0\left(\sup_{z\ge u}\frac1{L(z)^2}+\int_u^\infty\frac{dz}{zL(z)^2}\right).
\]
Moreover, Proposition~\ref{prop:radial-scalar}, $r\les v$, and $\la\asymp1$ imply
\[
\sup_{v>u}|r\pr_v\phi(u,v)|\les\epsilon_0\sup_{z\ge u}L(z)^{-1}\to0.
\]
The geometric bound also tends to zero. Since $r(u,v)=\int_u^v\la(u,v')dv'$, the relative radius error is controlled by the same bound for $\la-\frac12$. This proves \eqref{eq:geometry-limit}. For $s>1$, we have from \eqref{choiceLs>1}
\[
 \int_u^\infty\frac{dz}{zL(z)^2}
 =\frac{u^{-\min\{s-1,1/2\}}}{\min\{s-1,1/2\}}.
\]
The geometric and relative-radius errors are therefore $O(\epsilon_0u^{-\min\{s-1,1/2\}})$. The scalar derivative bound above becomes
\[
\sup_{v>u}|r\pr_v\phi(u,v)|\les\eps_0u^{-\min\left\{\frac{s-1}{2},\frac{1}{4}\right\}}.
\]
Together these give the rate in Theorem~\ref{thm:s=1}.
\section{Half-derivative stability at decay and regularity thresholds}
\label{half:sec:main-estimates}
For Theorem~\ref{thm:half-stability}, we take smooth data in the normalization~\eqref{eq:data} with
\begin{equation}\label{half:eq:small-data}
\|\Phi\|_{X^{1,1/2}}=\frac1{\sqrt2}[\Phi]_{\dot H^{1/2}(1,\infty)}\le\eps_0.
\end{equation}
\subsection{Fractional estimates}\label{half:sec:radial-averaging}
For the energy and source estimates, we use the following seminorm and pairing on an interval $I$.
\begin{equation}\label{half:eq:gagliardo-pairing}
 \begin{aligned}
 \left[f\right]_{\dot H^{1/2}(I)}^2
 &:=\int_I\int_I
 \frac{|f(x)-f(y)|^2}{|x-y|^2}dxdy,\\
 \langle f,g\rangle_{\dot H^{1/2}(I)}
 &:=\int_I\int_I
 \frac{(f(x)-f(y))(g(x)-g(y))}{|x-y|^2}dxdy.
 \end{aligned}
\end{equation}
\begin{lem}\label{half:lem:analytic-estimates}\label{half:lem:radial-averaging}
\begin{enumerate}
\renewcommand{\labelenumi}{\textup{(\roman{enumi})}}
\item Let $I=(r_0,r_1)$ be bounded,
$f\in H^{1/2}(I)$, $c\in\mathbb R$, and
\[
 \overline f(r)=\frac1{r-r_0}\int_{r_0}^rf(\rho)d\rho.
\]
Then
\begin{equation}\label{half:eq:averaging-bounds}
 \begin{split}
 [f-\overline f]_{\dot H^{1/2}(I)}
 &\le2[f]_{\dot H^{1/2}(I)},\\
 \int_I\frac{|f-\overline f|^2}{r-r_0}dr
 &\le\frac12[f]_{\dot H^{1/2}(I)}^2,\\
 \int_I\frac{|f-\overline f|^2}{(r-r_0)^2}dr
 &\le\frac{25}{9}\int_I\frac{|f-c|^2}{(r-r_0)^2}dr.
 \end{split}
\end{equation}
The last inequality holds whenever its right-hand side is finite.

\item For a bounded interval $J$, $f\in H^{1/2}(J)$,
and $g\in W^{1,2}(J)$,
\begin{equation}\label{half:eq:local-product}
 [gf]_{\dot H^{1/2}(J)}^2
 \les\|g\|_{L^\infty(J)}^2[f]_{\dot H^{1/2}(J)}^2
 +\|g\|_{L^\infty(J)}\|\pr_rg\|_{L^2(J)}
       \|f\|_{L^4(J)}^2,
\end{equation}
and
\begin{equation}\label{half:eq:interval-interpolation}
 \|f\|_{L^4(J)}^4
 \les\|f\|_{L^2(J)}^2
 \left(
 [f]_{\dot H^{1/2}(J)}^2
 +|J|^{-1}\|f\|_{L^2(J)}^2
 \right).
\end{equation}

\item Let $I=(0,R)$, $0<R<\infty$, and
\[
 J_j=(2^{j-1},\min\{2^{j+2},R\}),
 \qquad j\in\mathbb Z,\quad 2^j<R.
\]
For $f\in H^{1/2}(I)$ with $f/\sqrt r\in L^2(I)$,
\begin{equation}\label{half:eq:dyadic-localization}
 \begin{split}
 [f]_{\dot H^{1/2}(I)}^2
 +\int_I\frac{|f|^2}{r}dr
 \asymp\sum_{2^j<R}\left(
 [f]_{\dot H^{1/2}(J_j)}^2
 +|J_j|^{-1}\|f\|_{L^2(J_j)}^2
 \right).
 \end{split}
\end{equation}
\end{enumerate}
All implicit constants are universal.
\end{lem}

\begin{proof}
Part~\textup{(i)} follows from dilation invariance,
Minkowski's and Jensen's inequalities, and weighted
Hardy~\cite[Sections~6.13--6.14 and~9.9]{HLP}.
The product estimate in~\textup{(ii)} follows from the
product difference identity, H\"older's inequality, and
the $L^2$ maximal theorem~\cite[Chapter~I, Theorem~1(c)]{Stein1970},
applied to the zero extension of $\pr_rg$.
The interpolation estimate follows from fractional
Gagliardo--Nirenberg~\cite[Proposition~5.9]{MorosiPizzocchero},
using the extension on $(0,1)$
in~\cite[Lemmas~5.1--5.3 and proof of Theorem~5.4]{DNPV}
and rescaling.
The localization in~\textup{(iii)} follows from bounded
overlap and the decomposition into comparable and separated radii.
\end{proof}
\subsection{Energy and geometric bounds}
We use $G$ from \eqref{eq:G-definition} and $\Omega^2=-4\la\nu/(1-\mu)$. Regularity at the axis gives
\[
r(u,u)=m(u,u)=\mu(u,u)=0,\qquad\nu(u,u)=-\la(u,u),\qquad\pr_r(r\phi)(u,u)=\phi(u,u).
\]
Integration in $r$ yields
\[
\phi(u,r)=\frac1r\int_0^r\pr_r(r\phi)(u,r')dr',\qquad\pr_r(r\phi)-\phi=r\pr_r\phi.
\]
For $V>1$, on $\mathcal T_V=\{(u,v):1\le u\le v\le V\}$ and $u<V$ set
\begin{equation}\label{half:eq:difference-energy}
 \begin{aligned}
 E_V(u)&:=[\pr_r(r\phi)(u,\cdot)]_{\dot H^{1/2}(u,V)}^2,\\
 B_V(u)&:=\int_1^u\int_{u'}^V
 \frac{|\pr_r(r\phi)(u',v)-\pr_r(r\phi)(u',u')|^2}{(v-u')^2}dvdu'.
 \end{aligned}
\end{equation}
We have $E_V(1)\le[\Phi]_{\dot H^{1/2}(1,\infty)}^2\le2\eps_0^2$, $B_V(1)=0$, and
\begin{align*}
    B'_V(u)=\int_{u}^V\frac{|\pr_r(r\phi)(u,v)-\pr_r(r\phi)(u,u)|^2}{(v-u)^2}dv.
\end{align*}
We now control the initial data to obtain the geometric bounds needed for the bootstrap.
\begin{lem}\label{half:lem:initial-bounds}
For sufficiently small $\eps_0$, the following estimates hold on $C_1$:
\[
\la=\frac12,\qquad|\nu+\tfrac12|+\mu+|\Omega^2-1|\les\eps_0^2,\qquad\pr_r(r\phi)=\Phi.
\]
In particular, the initial cone is untrapped and $E_V(1)\le2\eps_0^2$.
\end{lem}
\begin{proof}
All quantities below are restricted to $C_1$, where $\la=1/2$, $r=(v-1)/2$, and $\pr_r(r\phi)=\Phi$. Since $\phi$ is the radial average of $\Phi(1+2r')$ on $C_1$, the second inequality in~\eqref{half:eq:averaging-bounds} and the change of variables $v=1+2r'$ give
\[
 \int_0^r r'|\pr_r\phi(1,r')|^2dr'\leq \frac12[\Phi]_{\dot H^{1/2}(1,1+2r)}^2\leq\|\Phi\|_{X^{1,1/2}}^2\leq \eps_0^2.
\]
The mass constraint \eqref{eq:mass-radial} implies
\[
 \pr_rm\le \pr_rm+r(\pr_r\phi)^2m=\frac12r^2(\pr_r\phi)^2,\qquad m(0)=0.
\]
Hence
\begin{align*}
 0\le\mu(r)\le\frac1r\int_0^r(r')^2|\pr_r\phi(r')|^2dr'
 \le\int_0^r r'|\pr_r\phi(r')|^2dr'
 \le\eps_0^2.
\end{align*}
Thus $1-\mu\ge1/2$. The initial radius constraint, with $\nu(0)=-1/2$, and \eqref{half:eq:metric-log} give
\[
 \frac{-2\nu(r)}{1-\mu(r)}
 =\Omega^2(r)
 =\exp\left(\int_0^r r'|\pr_r\phi(r')|^2dr'\right).
\]
The geometric bounds follow, while $E_V(1)\le2\eps_0^2$ follows directly from \eqref{half:eq:small-data}.
\end{proof}
Set $\eps=\eps_0^{2/3}$. On a finite characteristic triangle, assume the following bootstrap bounds
\begin{equation}\label{half:eq:bootstrap-bounds}
 \frac12-\la+\left|\nu+\frac12\right|+\mu\le\eps,\qquad E_V(u)+B_V(u)\le\eps^2. \tag{B.A.-Thm B}
\end{equation}
Lemma \ref{half:lem:initial-bounds} and continuity give an initial bootstrap interval. The metric bounds and integration of $dr/dv=\la$ from the axis give
\begin{equation}\label{half:eq:radial-comparison}
\frac{v-u}{4}\le r(u,v)\le v-u,\qquad\frac14\le\frac{dr}{dv}\le1.
\end{equation}
The half-derivative seminorms in $v\in(u,V)$ and $r\in(0,r(u,V))$ are equivalent with universal constants.
\begin{prop}\label{lanumuest-12}
Under the bootstrap assumptions \eqref{half:eq:bootstrap-bounds}, we have
\[
0\le\frac12-\la+\left|\nu+\frac12\right|+\mu+|\Omega^2-1|\les\eps_0,\qquad \|G(u,\cdot)\|_{L^\infty(u,V)}\les B_V'(u).
\]
\end{prop}
\begin{proof}
At fixed $u$, set $R=r(u,V)$ and $f(r)=\pr_r(r\phi)(u,r)$ on $(0,R)$. Then $\overline f=\phi$ and $f-\overline f=r\pr_r\phi$. The first two inequalities in~\eqref{half:eq:averaging-bounds}, and the third with $c=f(0)=\pr_r(r\phi)(u,u)$, yield, after the change of variables~\eqref{half:eq:radial-comparison},
 \begin{align}
 \begin{split}\label{half:eq:radial-energy-bounds}
 [r\pr_r\phi]_{\dot H^{1/2}(0,R)}^2+\int_0^Rr|\pr_r\phi|^2dr\les[f]_{\dot H^{1/2}(0,R)}^2&\les E_V(u),\\
 \int_0^R|\pr_r\phi|^2dr\les\int_0^R\frac{|f-f(0)|^2}{r^2}dr&\les B_V'(u).
 \end{split}
 \end{align} Integrating \eqref{half:eq:G-radial} from the regular axis, then using $r'\le r$ and the metric bootstrap, yields
\begin{equation}\label{half:eq:G-sup-bound}
 \begin{aligned}
 |G(u,r)|
 &=\frac1{r^2}\left|\int_0^r\frac{\nu}{1-\mu}(u,r')(r')^2|\pr_r\phi(u,r')|^2dr'\right|\\
 &\les\int_0^r|\pr_r\phi(u,r')|^2dr'\le\int_0^R|\pr_r\phi(u,r')|^2dr'\les B_V'(u).
 \end{aligned}
\end{equation}
The radial mass equation \eqref{eq:mass-radial} gives
\[
0\le\mu(u,r)\le\frac1r\int_0^r(r')^2|\pr_r\phi(u,r')|^2dr'\le\int_0^r r'|\pr_r\phi(u,r')|^2dr'\les E_V(u)\le\eps^2.
\]
Integrating $\pr_u\log\la=G$ at fixed $v$, using $\la(1,v)=1/2$, $G\le0$, and \eqref{half:eq:G-sup-bound}, we obtain
\[
 0\le-\log(2\la(u,v))
 =\int_1^u(-G)(u',v)du'
 \les B_V(u)\le\eps^2.
\]
Thus $0\le1/2-\la\les\eps^2$, including at the axis. At the axis, $\nu(u,u)=-\la(u,u)$ and $\mu(u,u)=0$. Integrating \eqref{half:eq:metric-log} therefore gives
\[
\frac{-\nu(u,r)}{1-\mu(u,r)}=\la(u,u)\exp\left(\int_0^r r'|\pr_r\phi(u,r')|^2dr'\right).
\]
The preceding bounds imply $|\nu+1/2|\les\eps^2$. Since $\Omega^2=-4\la\nu/(1-\mu)$, we also obtain $|\Omega^2-1|\les\eps^2$. Finally, $\eps^2=\eps_0^{4/3}\le\eps_0$ gives the stated geometric estimates.
\end{proof}
\subsection{The nonlinear source estimate}
\label{half:sec:source}
\begin{lem}
\label{half:lem:cubic-estimate}
Fix $u$ in a regular untrapped solution. Let $I=(0,R)$, where $0<R<\infty$. If $|\nu/(1-\mu)|\les1$, then
\begin{equation}\label{half:eq:source-interval}
 [rG\pr_r\phi]_{\dot H^{1/2}(I)}\les\left(\|G\|_{L^\infty(I)}+\int_I|\pr_r\phi|^2dr\right)\left([r\pr_r\phi]_{\dot H^{1/2}(I)}^2+\int_Ir|\pr_r\phi|^2dr\right)^{1/2}.
\end{equation}
Under~\eqref{half:eq:bootstrap-bounds},
\begin{equation}\label{half:eq:cubic-estimate}
 [(rG\pr_r\phi)(u,\cdot)]_{\dot H^{1/2}(u,V)}\les\eps B_V'(u).
\end{equation}
\end{lem}
\begin{proof}
At fixed $u$ write $f=r\pr_r\phi=\pr_r(r\phi)-\phi$. On $J\subset I$ with $r\asymp|J|$, use~\eqref{half:eq:G-radial} to obtain
\[
\pr_rG=\frac{\nu}{1-\mu}\frac{f^2}{r^2}-\frac{2G}{r}.
\]
Hence
\begin{align*}
 \|\pr_rG\|_{L^2(J)}
 &\les |J|^{-2}\|f\|_{L^4(J)}^2+|J|^{-1}\|G\|_{L^2(J)}\les |J|^{-2}\|f\|_{L^4(J)}^2+|J|^{-1/2}\|G\|_{L^\infty(I)}.
\end{align*}
Therefore, \eqref{half:eq:interval-interpolation} and $r\asymp|J|$ give
\begin{align*}
 |J|^{-2}\|f\|_{L^4(J)}^4
 &\les |J|^{-2}\|f\|_{L^2(J)}^2\left([f]_{\dot H^{1/2}(J)}^2+|J|^{-1}\|f\|_{L^2(J)}^2\right)\\
 &\les\left(\int_I|\pr_r\phi|^2dr\right)\left([f]_{\dot H^{1/2}(J)}^2+|J|^{-1}\|f\|_{L^2(J)}^2\right).
\end{align*}
Taking the square root of~\eqref{half:eq:interval-interpolation} gives
\begin{align*}
 |J|^{-1/2}\|f\|_{L^4(J)}^2
 &\les\bigl(|J|^{-1}\|f\|_{L^2(J)}^2\bigr)^{1/2}\left([f]_{\dot H^{1/2}(J)}^2+|J|^{-1}\|f\|_{L^2(J)}^2\right)^{1/2}\\
 &\le[f]_{\dot H^{1/2}(J)}^2+|J|^{-1}\|f\|_{L^2(J)}^2.
\end{align*}
Taking $g=G$ in~\eqref{half:eq:local-product} and using these bounds, we obtain
\begin{align*}
 [Gf]_{\dot H^{1/2}(J)}^2
 &\les\|G\|_{L^\infty(J)}^2[f]_{\dot H^{1/2}(J)}^2+\|G\|_{L^\infty(J)}\|\pr_rG\|_{L^2(J)}\|f\|_{L^4(J)}^2\\
 &\les \|G\|_{L^\infty(I)}^2[f]_{\dot H^{1/2}(J)}^2+\|G\|_{L^\infty(I)}|J|^{-2}\|f\|_{L^4(J)}^4+\|G\|_{L^\infty(I)}^2|J|^{-1/2}\|f\|_{L^4(J)}^2\\
 &\les\|G\|_{L^\infty(I)}^2[f]_{\dot H^{1/2}(J)}^2+\|G\|_{L^\infty(I)}\left(\int_I|\pr_r\phi|^2dr\right)\left([f]_{\dot H^{1/2}(J)}^2+|J|^{-1}\|f\|_{L^2(J)}^2\right)\\
 &+\|G\|_{L^\infty(I)}^2\left([f]_{\dot H^{1/2}(J)}^2+|J|^{-1}\|f\|_{L^2(J)}^2\right)\\
 &\les\|G\|_{L^\infty(I)}\left(\|G\|_{L^\infty(I)}+\int_I|\pr_r\phi|^2dr\right)\left([f]_{\dot H^{1/2}(J)}^2+|J|^{-1}\|f\|_{L^2(J)}^2\right).
\end{align*}
For the intervals $J_j$ in Lemma~\ref{half:lem:analytic-estimates}\textup{(iii)}, applying~\eqref{half:eq:dyadic-localization} to $Gf$ and then to $f$ gives
\begin{align*}
 [Gf]_{\dot H^{1/2}(I)}^2
 &\les\sum_{2^j<R}\left([Gf]_{\dot H^{1/2}(J_j)}^2+|J_j|^{-1}\|Gf\|_{L^2(J_j)}^2\right)\\
 &\les\|G\|_{L^\infty(I)}\left(\|G\|_{L^\infty(I)}+\int_I|\pr_r\phi|^2dr\right)\sum_{2^j<R}\left([f]_{\dot H^{1/2}(J_j)}^2+|J_j|^{-1}\|f\|_{L^2(J_j)}^2\right)\\
 &\les\|G\|_{L^\infty(I)}\left(\|G\|_{L^\infty(I)}+\int_I|\pr_r\phi|^2dr\right)\left([f]_{\dot H^{1/2}(I)}^2+\int_I\frac{f^2}{r}dr\right).
\end{align*}
Taking a square root proves~\eqref{half:eq:source-interval}. Under~\eqref{half:eq:bootstrap-bounds}, take $I=(0,r(u,V))$. By~\eqref{half:eq:radial-energy-bounds}--\eqref{half:eq:G-sup-bound}, we have already proven
\begin{align*}
    \|G\|_{L^\infty(I)}+\int_I|\pr_r\phi|^2dr &\les B_V'(u),\\
    [r\pr_r\phi]_{\dot H^{1/2}(I)}^2+\int_Ir|\pr_r\phi|^2dr &\les E_V(u)\le\eps^2.
\end{align*}
Since $\la\asymp1$, \eqref{half:eq:radial-comparison} gives
\[
[(rG\pr_r\phi)(u,\cdot)]_{\dot H^{1/2}(u,V)}\les[rG\pr_r\phi]_{\dot H^{1/2}(I)}\les B_V^{\prime}(u)E_V(u)^{1/2}\le\eps B_V^{\prime}(u),
\]
which proves~\eqref{half:eq:cubic-estimate}.
\end{proof}
\subsection{Proof of Theorem~\ref{thm:half-stability}}\label{half:sec:proof}
\noindent\textbf{Step 1. Energy estimates and bootstrap closure.}
Differentiating the two lower endpoints in $E_V$ contributes $-2B_V'(u)$. Each contribution is $-B_V'(u)$, since one of the variables in the double integral is set equal to $u$. Using \eqref{half:eq:h-transport}, \eqref{half:eq:radial-energy-bounds}, and Lemma~\ref{half:lem:cubic-estimate}, we obtain
\begin{align*}
 E_V'(u)+2B_V'(u)=-2\langle\pr_r(r\phi),rG\pr_r\phi\rangle_{\dot H^{1/2}(u,V)}\le2\sqrt{E_V(u)}[rG\pr_r\phi]_{\dot H^{1/2}(u,V)}\les\eps^2B_V'(u).
\end{align*}
For small enough $\eps_0$, we absorb the term $\eps^2B_V'(u)$ and integrate from $1$ to get
\begin{equation}\label{half:eq:closed-energy}
 E_V'(u)+B_V'(u)\le0,\qquad E_V(u)+B_V(u)\le E_V(1)\le2\eps_0^2.
\end{equation}
This bound and Proposition~\ref{lanumuest-12} strictly improve~\eqref{half:eq:bootstrap-bounds} uniformly in $V$.

\noindent\textbf{Step 2. Global existence and completeness.}
Fix $V>1$. For $1\le u<V$, integrate radially from the axis on $C_u$ to get
\[
 \|\pr_r\phi(u,\cdot)\|_{L^\infty(u,V)}
 \les\|\pr_r^2(r\phi)(u,\cdot)\|_{L^\infty(u,V)}.
\]
Integrate the derivative of $|r\pr_r\phi|^3$ from its zero value at the axis and use the preceding bound.
\begin{align*}
 |r\pr_r\phi(u,r)|^3\les\int_0^r\rho^2|\pr_\rho\phi|^2
       \left(|\pr_\rho^2(\rho\phi)|+|\pr_\rho\phi|\right)d\rho\les r^2\|\pr_r^2(r\phi)(u,\cdot)\|_{L^\infty(u,V)}
       \int_0^r|\pr_\rho\phi|^2d\rho.
\end{align*}
In~\eqref{eq:pruprrrphi} use~\eqref{half:eq:G-sup-bound} for the $G$ term. For the cubic, divide by $r^2$ in the preceding estimate and apply~\eqref{half:eq:radial-energy-bounds}. This gives
\[
 \left\|\pr_u\bigl(\la^2\pr_r^2(r\phi)\bigr)(u,\cdot)\right\|_{L^\infty(u,V)}
 \les B_V'(u)\left\|\bigl(\la^2\pr_r^2(r\phi)\bigr)(u,\cdot)\right\|_{L^\infty(u,V)}.
\]
Integrate at fixed $v$ before taking the supremum over $v\in(u,V)$. Gr\"onwall then bounds $\|\bigl(\la^2\pr_r^2(r\phi)\bigr)(u,\cdot)\|_{L^\infty(u,V)}$ by $\|\Phi'/2\|_{L^\infty(1,V)}e^{CB_V(u)}$. Since $\la\asymp1$, this bounds $\pr_r^2(r\phi)$ in $L^\infty(u,V)$. Transport and radial averaging bound $\pr_r(r\phi)$ and $\phi$ in the same norm. Existence of a unique global smooth solution follows from Proposition~\ref{app:prop:continuation}.

Use Cauchy--Schwarz in~\eqref{eq:incoming-scalar-integral} and then~\eqref{half:eq:radial-energy-bounds} to get
\[
r|\pr_u\phi(u,v)|^2\les\int_0^r|\pr_r\phi(u,r')|^2dr'\les B_V'(u),\qquad u\le v\le V.
\]
On a compact geodesic segment take $V>\max v$. The energy argument in \hyperref[step:causal-completeness]{Step~2} of Section~\ref{sec:proofA} now uses $B_V'(u)$ in place of $\eps_0^2/(uL(u)^2)$ and yields
$$
\mathcal E(\tau)\le\mathcal E(0)e^{C[B_V(u(\tau))-B_V(u(0))]}.
$$
By~\eqref{half:eq:closed-energy}, $\mathcal E(\tau)\les\mathcal E(0)$ uniformly in $V$. The inequality $\dot u+\dot v+|\dot r|+r|\dot\omega|_{\mathbb S^2}\les\mathcal E$ bounds the velocity in regular coordinates and gives future causal geodesic completeness.

\noindent\textbf{Step 3. Energy decay and geometric convergence.}
Write $E_\infty(u)=\lim_{V\to\infty}E_V(u)$ and $B_\infty(u)=\lim_{V\to\infty}B_V(u)$. Passing to the limit in~\eqref{half:eq:closed-energy} proves~\eqref{half:eq:main-energy} and gives $\sup_{u\ge1}B_\infty(u)\le2\eps_0^2$. Monotone convergence gives $B_\infty(u)=\int_1^u\lim_{V\to\infty}B_V'(u')du'$. In particular, $B_\infty$ is absolutely continuous and $B_\infty'(u)=\lim_{V\to\infty}B_V'(u)$ almost everywhere.

Integrating \eqref{half:eq:h-transport} and applying Minkowski's inequality gives
\[
 \sqrt{E_\infty(u)}
 \le[\Phi]_{\dot H^{1/2}(u,\infty)}
 +\int_1^u
 [(rG\pr_r\phi)(u',\cdot)]_{\dot H^{1/2}(u,\infty)}
 du'.
\]
The initial seminorm tail tends to zero when $u\to\infty$. The source seminorm tail also tends to zero for a.e. $u'$, while Lemma~\ref{half:lem:cubic-estimate} bounds it by the integrable function $\les\eps B_\infty'(u')$. Dominated convergence theorem therefore proves \eqref{half:eq:energy-decay}.

The mass estimate gives $\mu\les E_\infty(u)\to0$. Moreover, \eqref{eq:G-definition} and \eqref{half:eq:G-sup-bound} give
\begin{equation}\label{estG}
|G(u,v)|\les\frac{E_\infty(u)}{r(u,v)},\qquad\|G(u,\cdot)\|_{L^\infty(u,\infty)}\les B_\infty'(u)\quad{\text{for a.e. }u}.
\end{equation}
For $v\ge u>U>1$, split~\eqref{eq:lambda-u} at $U$ and apply~\eqref{estG} with $r(u',v)\ge(v-u')/4$ to get
\begin{align*}
 -\log(2\la(u,v))
 &\les\int_1^U\frac{E_\infty(u')}{r(u',v)}du'+B_\infty(u)-B_\infty(U)\\
 &\les\frac1{u-U}\int_1^UE_\infty(u')du'+\sup_{u'\ge U}B_\infty(u')-B_\infty(U)\\
 &\les\frac{\eps_0^2(U-1)}{u-U}+\sup_{u'\ge U}B_\infty(u')-B_\infty(U).
\end{align*}
Since $B_\infty$ is bounded and increasing, choose $U$ to make its tail small. The first term then tends to zero as $u\to\infty$, uniformly in $v\ge u$, giving
\[
 \sup_{v\ge u}|\la(u,v)-1/2|\to 0.
\]
Integrate~\eqref{half:eq:metric-log} from the axis, where $\nu(u,u)=-\la(u,u)$. The radial integral is $O(E_\infty(u))$ by~\eqref{half:eq:radial-energy-bounds}. Together with $\mu\to0$, this gives $\nu\to-1/2$ uniformly and then $\Omega^2\to1$ by~\eqref{eq:geometry-definitions}. Integrating $\la$ from the axis bounds the relative radius error by $2\sup_{v\ge u}|\la-1/2|$. This proves~\eqref{eq:geometry-limit-12}. The inclusion $X^{s,q}\subseteq X^{1,1/2}$ covers $s\ge1$, $q\ge1/2$.
\section{Trapped-sphere formation below the thresholds}\label{sec:instability}
Choose $\psi\in C_c^\infty((0,1))$ with $0\le\psi\le1$, equal to $1$ on $[1/4,3/4]$, and set
\begin{equation}\label{eq:seed-size}
 \sigma_N=e^{-N},\qquad
 \Phi_N(v):=10\sqrt N\,\psi\left(\frac{v-2}{\sigma_N}\right).
\end{equation}
Thus $\Phi_N\in C_c^\infty((2,2+\si_N))$ and $\|\Phi_N\|_{L^2(1,\infty)}^2\le100N\sigma_N\to 0$. The scalar data give
\begin{equation}\label{initial:scalar}
\phi(1,v)=\frac1{v-1}\int_1^v\Phi_N(v')dv',\qquad (r\pr_r\phi)(1,v)=\Phi_N(v)-\phi(1,v).
\end{equation}
For the initial data \eqref{eq:seed-size}, the small width of the support gives an initially untrapped cone with the following geometric bounds.
\begin{prop}\label{initial:prop}
For sufficiently large $N$, the initial cone is smooth and every sphere $v>1$ is strictly untrapped. Furthermore, we have the following control of the geometry
\begin{equation}\label{eq:seed-geometry}
 \sup_{v\ge1}\left(\left|\nu+\frac12\right|+\mu+|\Omega^2-1|\right)(1,v)
 \les N\sigma_N,
 \qquad \nu(1,v)\le-\frac12,\qquad 0\le\mu(1,v)<\frac12.
\end{equation}
\end{prop}
\begin{proof}
Throughout this proof, all quantities are restricted to the initial cone $C_1$, where $\la=1/2$, $r=(v-1)/2$, and $\Omega^2=-2\nu/(1-\mu)$. Using \eqref{half:eq:metric-log} and \eqref{initial:scalar}, we compute
\[
\pr_v\log\Om^2=\la r|\pr_r\phi|^2=\frac{|\Phi_N(v)-\phi(1,v)|^2}{v-1}.
\]
Integrate from the vertex using $\Omega^2(1,1)=1$
\begin{equation}\label{logOm2}
 \log\Omega^2(1,v)
 =\int_1^v\frac{|\Phi_N(v')-\phi(1,v')|^2}{v'-1}dv'.
\end{equation}
Thus $\Omega^2(1,v)\ge1$ is nondecreasing, and~\eqref{eq:nu-radial} and~\eqref{eq:G-definition} give
\[
\pr_v\bigl((v-1)\nu\bigr)=\nu+(v-1)\la G=\nu+rG=\frac{\nu}{1-\mu}=-\frac{\Omega^2}{2}.
\]
Integration from $v=1$ and $\Omega^2=-2\nu/(1-\mu)$ give
\[
-2\nu(1,v)=\frac1{v-1}\int_1^v\Omega^2(1,v')dv',\qquad 1-\mu(1,v)=\frac{\int_1^v\Omega^2(1,v')dv'}{(v-1)\Omega^2(1,v)}.
\]
The positivity and monotonicity of $\Om^2$ imply
\[
 \nu(1,v)\le-\frac12,\qquad 0\le\mu(1,v)<1.
\]
Since the data vanish for $v\le2$, these identities give Minkowski geometry there. It therefore suffices to estimate $v\ge2$. First, \eqref{initial:scalar} gives
\begin{equation}\label{v-1phi2}
\pr_v\phi(1,v)=\frac{\Phi_N(v)-\phi(1,v)}{v-1},\qquad\pr_v\bigl((v-1)\phi(1,v)^2\bigr)=2\phi(1,v)\Phi_N(v)-\phi(1,v)^2.
\end{equation}
Consequently,
\[
\int_1^v|\Phi_N(v')-\phi(1,v')|^2dv'=\int_1^v|\Phi_N(v')|^2dv'-(v-1)|\phi(1,v)|^2.
\]
Integrating \eqref{eq:mass-radial}, with $m(1,1)=0$, and using $0\le1-\mu\le1$, we obtain
\begin{align*}
 0\le\mu(1,v)
 &=\frac1{v-1}\int_1^v
 (1-\mu(1,v'))|\Phi_N(v')-\phi(1,v')|^2dv'\\
 &\le\frac1{v-1}\int_1^v
 |\Phi_N(v')-\phi(1,v')|^2dv'\\
 &\le\frac1{v-1}\int_2^{2+\sigma_N}|\Phi_N(v')|^2dv'\le100N\sigma_N.
\end{align*}
The vertex term vanishes since $\phi=0$ near $v=1$. Using~\eqref{v-1phi2}, integrate~\eqref{logOm2} by parts to get
\begin{align*}
 0\le\log\Omega^2(1,v)
 &=\int_1^v\frac{|\Phi_N(v')|^2-|\phi(1,v')|^2}{v'-1}dv'-|\phi(1,v)|^2\\
 &\le\int_2^{2+\sigma_N}\frac{|\Phi_N(v')|^2}{v'-1}dv'\\
 &\le\int_2^{2+\sigma_N}|\Phi_N(v')|^2dv'\le100N\sigma_N,
\end{align*}
so exponentiating for large $N$ gives
\[
0\le\Om^2(1,v)-1\le2\log\Om^2(1,v)\le200N\sigma_N.
\]
Averaging $\Omega^2-1$ in the formula for $\nu$ gives
\begin{align*}
0\leq-\nu(1,v)-\frac12=\frac1{2(v-1)}\int_1^v\bigl(\Omega^2(1,v')-1\bigr)dv'\leq\frac12\sup_{v'\ge1}\bigl(\Omega^2(1,v')-1\bigr)\leq 100N\sigma_N.
\end{align*}
Together with the mass bound this gives
\[
\sup_{v\ge1}\left(|\nu+\tfrac12|+\mu+|\Omega^2-1|\right)(1,v)\les N\sigma_N,
\]
which is~\eqref{eq:seed-geometry}. The constraints give smooth fields, with Minkowski data near the vertex. For large $N$, $\la=1/2$, $\nu<0$, and $1-\mu>1/2$, so the spheres $v>1$ are strictly untrapped.
\end{proof}
\begin{prop}\label{trapping:prop}
For sufficiently large $N$, the development of the initial data \eqref{eq:seed-size} contains a strictly trapped sphere at $u>1$. Its maximal smooth future development is future null geodesically incomplete.
\end{prop}
\begin{proof}
On $2+\sigma_N/4\le v\le2+3\si_N/4$, we have $\Phi_N(v)=10\sqrt N$. By \eqref{initial:scalar},
\[
0\le\phi(1,v)=\frac1{v-1}\int_2^v\Phi_N(v')dv'\le\frac{10\sqrt N(v-2)}{v-1}\le10\sqrt N\sigma_N.
\]
Therefore, for $N$ sufficiently large
\[
(r\pr_r\phi)(1,v)=\Phi_N(v)-\phi(1,v)\geq 10\sqrt N(1-\sigma_N)\geq 5\sqrt N.
\]
Proposition~\ref{initial:prop} also gives $1-\mu\ge1/2$. Integrating \eqref{eq:mass-radial} along the initial cone yields
\begin{equation}\label{trapping:mass}
\begin{aligned}
\frac{2[m(1,2+\si_N)-m(1,2)]}{r(1,2+\si_N)}&=\frac1{1+\si_N}\int_2^{2+\si_N}(1-\mu)|r\pr_r\phi|^2(1,v)dv\\
&\geq\frac{25N}{2(1+\si_N)}\int_{2+\si_N/4}^{2+3\si_N/4}dv\\
&=\frac{25N\si_N}{4(1+\si_N)}>\frac92\si_N.
\end{aligned}
\end{equation}
The incoming side $v=2$ is Minkowskian with $\phi=0$ by uniqueness, since $\Phi_N=0$ for $v\le2$. We use An--Lim's mass concentration criterion in the normalization of their appendix proof~\cite[Theorem~1.2]{AnLim}, with scalar field $\phi/\sqrt{4\pi}$. The width on $C_1$ is $\si_N/2$ and the inner radius is $1/2$, so the relative width is $\si_N$. The normalized mass input is the left side of~\eqref{trapping:mass}, which exceeds the required $9\si_N/2$.

If no trapped sphere forms before $u_1=2-\si_N/2$, then $\la\geq 0$ since $\nu<0$ by Raychaudhuri. Thus $r(u,v)\geq r(u_1,2)=\si_N/4$ whenever the solution is defined in $[1,u_1]\times [2,2+\si_N]$, and Dafermos's extension principle~\cite[Theorem~3.1]{Dafermos} gives continuation to $u_1$. Since $m\ge m(u,2)=0$, we have $\pr_u\la=-\Om^2m/(2r^2)\leq 0$, so $\la(u_1,v)\leq 1/2$. Therefore,
\begin{align*}
r(u_1,2+\si_N)-r(u_1,2)=\int_2^{2+\si_N}\la(u_1,v)\,dv\le\frac{\si_N}{2},
\end{align*}
hence $r(u_1,2+\si_N)\le3\si_N/4$. Since $r(1,2+\si_N)=(1+\si_N)/2$, the target outer radius in An--Lim's criterion is $3\si_N/2$. By continuity, there exists $u_*\in(1,u_1)$ such that $r(u_*,2+\si_N)=3\si_N/2$.

By An--Lim's criterion~\cite[Theorem~1.2]{AnLim}, there is a sphere with $\la\leq 0$ in $[1,u_*]\times [2,2+\si_N]$. Since $\nu<0$, this sphere is strictly trapped if $\la<0$. If $\la=0$, then $2m/r=1$ and $\pr_u\la=-\Om^2/(4r)<0$. Increasing $u$ slightly gives a strictly trapped sphere before $u_1$. In either case, this contradicts our assumption. Penrose's argument~\cite{Penrose1965} then gives future null geodesic incompleteness.
\end{proof}
\begin{proof}[Proof of Theorem~\ref{thm:instability}]
\noindent\textbf{The case $q<1/2$.}
For $0<q<1/2$, take $\Phi(v)=\psi(v-2)$ and $A=\si_N$ in Proposition~\ref{prop:scaling}\textup{(ii)}.
\begin{align*}
\|\Phi_N\|_{X^{s,q}}^2&=100N\left\|\psi\!\left(\frac{\cdot-2}{\si_N}\right)\right\|_{X^{s,q}}^2=100N\left\|\Phi_{\si_N}(\cdot-1+\si_N)\right\|_{X^{s,q}}^2\\
&\les_{s,q}N\|\Phi_{\si_N}\|_{X^{s,q}}^2\les_{s,q,\psi}Ne^{-(1-2q)N}\to 0.
\end{align*}
For $q=0$, use~\eqref{eq:X-norm}. The remaining conclusions follow from Propositions~\ref{initial:prop} and~\ref{trapping:prop}.

\noindent\textbf{The case $s<1$.}
Take $\max\{s,1/2\}<s'<1$, $A_N=e^{N^2}$, and an integer $M\ge\max\{1,q\}$. By monotonicity in $s$ and Proposition~\ref{prop:scaling}\textup{(i)},
\[
\|\Phi_{N,A_N}\|_{X^{s,q}}^2\les_{s',M}A_N^{s'-1}\|\Phi_N\|_{C^M}^2\les_{s',M}Ne^{2MN+(s'-1)N^2}\to0.
\]
Dilation preserves the remaining conclusions.
\end{proof}
\section{Geometry at future null infinity}\label{sec:null-infinity}
Constants in this section may depend on the additional finite initial norms.
\subsection{A continuous null boundary}
\begin{proof}[Proof of Theorem~\ref{thm:null-infinity}\textup{(i)}]
Since $x=1/r$,
\[
\left.\pr_x\right|_u=-r^2\pr_r=-\la^{-1}r^2\pr_v,\qquad
\left.\pr_u\right|_x=\left.\pr_u\right|_v-\la^{-1}\nu\pr_v.
\]
Fix $A>1$. Letting $V\to\infty$ in the bound by $E_V(u)^{1/2}B_V'(u)$ in the proof of Lemma~\ref{half:lem:cubic-estimate} and using~\eqref{half:eq:main-energy}, we obtain
\[
 [(rG\pr_r\phi)(u,\cdot)]_{\dot H^{1/2}(u,\infty)}\les\eps_0B_\infty'(u)
 \quad\text{for a.e. }u,\qquad \sup_{u\geq 1}B_\infty(u)\les\eps_0^2.
\]
For $V>A$, Minkowski's inequality,~\eqref{half:eq:h-transport}, and dominated convergence yield
\begin{align*}
 &\sup_{1\le u\le A}\left(\int_V^\infty\int_u^v
 \frac{|\pr_r(r\phi)(u,v)-\pr_r(r\phi)(u,w)|^2}{(v-w)^2}d  wdv\right)^{1/2}\\
 &\quad\le\left(\int_V^\infty\int_1^v
 \frac{|\Phi(v)-\Phi(w)|^2}{(v-w)^2}dwdv\right)^{1/2}\\
 &\quad+\int_1^A\left(\int_V^\infty\int_{u'}^v
 \frac{|(rG\pr_r\phi)(u',v)-(rG\pr_r\phi)(u',w)|^2}{(v-w)^2}dwdv\right)^{1/2}du'
 \to0,\qquad (V\to \infty).
\end{align*}
Radial averaging and~\eqref{half:eq:radial-comparison} give
\begin{align*}
\int_R^\infty r|\pr_r\phi(u,r)|^2dr&\le\int_R^\infty\int_0^r\frac{|\pr_r(r\phi)(u,r)-\pr_r(r\phi)(u,\rho)|^2}{r^2}d\rho dr\\
&\les\int_{u+R}^\infty\int_u^v\frac{|\pr_r(r\phi)(u,v)-\pr_r(r\phi)(u,w)|^2}{(v-w)^2}dwdv\to0,\qquad(R\to \infty)
\end{align*}
uniformly for $1\le u\le A$. Thus
\begin{align*}
0\le\mu(u,r)\le\frac1r\int_0^r\rho^2|\pr_\rho\phi(u,\rho)|^2d\rho\le\frac{R}{r}\int_0^R\rho|\pr_\rho\phi(u,\rho)|^2d\rho+\int_R^\infty\rho|\pr_\rho\phi(u,\rho)|^2d\rho.
\end{align*}
Thus $\mu(u,r)\to0$ locally uniformly in $u$, and~\eqref{half:eq:metric-log} gives
\begin{equation}\label{null:eq:nu-limit}
\nu_\infty(u):=\lim_{v\to\infty}\nu(u,v)=-\la(u,u)\exp\left(\int_0^\infty r|\pr_r\phi(u,r)|^2dr\right),\qquad-\nu_\infty(u)\asymp1.
\end{equation}
The uniform tails give continuity of $\nu_\infty$. By~\eqref{eq:lambda-u} and~\eqref{estG}, for $v\ge2A$,
\[
|\log(2\la(u,v))|\le\int_1^u|G(u',v)|du'\les\int_1^u\frac{E_\infty(u')}{v-u'}du'\les_A v^{-1}\to0.
\]
Using $\Omega^2=-4\la\nu/(1-\mu)$, we obtain, locally uniformly in $u$,
\begin{equation}\label{null:eq:metric-limits}
 \la(u,v)\to\frac12,\qquad \mu(u,v)\to0,\qquad
 \nu(u,v)\to\nu_\infty(u)<0,\qquad \Omega^2(u,v)\to-2\nu_\infty(u).
\end{equation}
Thus,
\begin{equation}\label{null:eq:conformal-metric}
r^{-2}\g=\frac{\Omega^2}{\la}dudx+x^2\frac{\Omega^2\nu}{\la}du^2+d\omega^2,\qquad
 \frac{\Omega^2}{\la}=\frac{-4\nu}{1-\mu}\to-4\nu_\infty(u).
\end{equation}
Since $U'=-2\nu_\infty\asymp1$, we have $U(u)\to\infty$ and
\[
\lim_{x\downarrow0}r^{-2}\g=-4\nu_\infty(u)dudx+d\omega^2=2dUdx+d\omega^2.
\]
Moreover,
\[
\left.(x^2\g)\right|_{T\mathcal I^+}=d\omega^2,\qquad(x^2\g)^{-1}(dx,dx)=(1-\mu)x^2\to0.
\]
Along an outgoing radial null ray with affine parameter $\tau$,
\[
\frac{d}{dv}\left(\Omega^{-2}\frac{d\tau}{dv}\right)=0,\qquad\frac{d\tau}{dv}\asymp\Omega^2\asymp1,\qquad \tau\to\infty.
\]
The ray $(u_0,v,\omega_0)$ from $(u_0,v_0,\omega_0)$ reaches $(u_0,0,\omega_0)\in\mathcal I^+$. The black-hole region, the complement of this causal past, is therefore empty.
\end{proof}
\subsection{Scalar limits, mass, and radiation}\label{null:subsec:mass-radiation}
\begin{prop}\label{null:prop:mass-radiation}
Let $(r,\phi,m)$ be the characteristic development of Theorem~\ref{thm:half-stability}, and let $c\in\mathbb R$.
\begin{enumerate}
\item[(i)] If $\Phi(v)\to c$, then $\pr_r(r\phi)(u,v)\to c$ and $\phi(u,v)\to c$, locally uniformly in $u$ as $v\to\infty$.
\item[(ii)] If $\Phi-c\in L^2(1,\infty)$, then
\begin{equation}\label{null:eq:bondi-mass}
 M_B(u)<\infty,\qquad M_B(u)\downarrow0\quad(u\to\infty).
\end{equation}
The limit $\lim_{v\to\infty}r\pr_u\phi(u,v)$ exists locally uniformly in $u$, $M_B\in C^1$, and~\eqref{null:eq:mass-loss}--\eqref{null:eq:total-flux} hold.

\item[(iii)] If $\Phi-c\in L^1(1,\infty)\cap L^2(1,\infty)$, then $\lim_{v\to\infty}r(\phi-c)(u,v)$ exists locally uniformly in $u$. This limit is $C^1$ in $u$, with derivative $\lim_{v\to\infty}r\pr_u\phi$.
\end{enumerate}
\end{prop}
\begin{proof}
For $\Phi-c\in L^\infty$, radial averaging and~\eqref{half:eq:h-transport} give
\begin{equation}\label{null:eq:bounded-profile}
 \|\pr_r(r\phi)(u,\cdot)-c\|_{L^\infty(u,\infty)}\leq\|\Phi-c\|_{L^\infty(1,\infty)}e^{CB_\infty(u)}.
\end{equation}
Still assuming $\Phi-c\in L^\infty(1,\infty)$, fix $A>1$.
For $1\le u\le A$ and $v\ge2A$,
\begin{equation}\label{null:eq:profile-transport}
 |G(u,v)|\les\frac{E_\infty(u)}{r(u,v)}\les_A v^{-1},\qquad
 |\pr_r(r\phi)(u,v)-\Phi(v)|
 \le\int_1^u|rG\pr_r\phi|(u',v)du'\les_A v^{-1}.
\end{equation}
If $\Phi(v)\to c$, this boundedness assumption holds automatically.
The comparison above and radial averaging therefore prove
\textup{(i)}.

For $\Phi-c\in L^2$, Hardy's inequality gives
\begin{equation}\label{null:eq:L2-averaging}
\|\phi-c\|_{L^2(dr)}\le2\|\pr_r(r\phi)-c\|_{L^2(dr)},\qquad\|r\pr_r\phi\|_{L^2(dr)}\le3\|\pr_r(r\phi)-c\|_{L^2(dr)}.
\end{equation}
The same bounds hold on finite intervals. Since $dr=\la dv\asymp dv$,~\eqref{half:eq:h-transport} and~\eqref{half:eq:G-sup-bound} imply
\begin{align*}
\frac{d}{du}\int_u^V|\pr_r(r\phi)-c|^2dv&=-|\phi(u,u)-c|^2-2\int_u^V(\pr_r(r\phi)-c)rG\pr_r\phi dv\\
&\les B_V'(u)\int_u^V|\pr_r(r\phi)-c|^2dv.
\end{align*}
By~\eqref{half:eq:main-energy} and Gr\"onwall's inequality,
\begin{align*}
 \sup_{u\ge1}\|\pr_r(r\phi)(u,\cdot)-c\|_{L^2(u,\infty)}
 &\les e^{C\eps_0^2}\|\Phi-c\|_{L^2(1,\infty)},\\
 \int_1^\infty\|(rG\pr_r\phi)(u,\cdot)\|_{L^2(u,\infty)}du
 &\les\|\Phi-c\|_{L^2(1,\infty)}\lim_{u\to\infty} B_\infty(u)<\infty.
\end{align*}
Transport and dominated convergence yield
\[
\|\pr_r(r\phi)(u,\cdot)-c\|_{L^2(u,\infty)}\le\|\Phi-c\|_{L^2(u,\infty)}+\int_1^u\|(rG\pr_r\phi)(u',\cdot)\|_{L^2(u,\infty)}du'\to0.
\]
Consequently,
\[
 0\le m(u,r)\le\frac12\int_0^r|\rho\pr_\rho\phi|^2d\rho
 \les\|\pr_r(r\phi)(u,\cdot)-c\|_{L^2(u,\infty)}^2,\qquad
 \pr_rm\ge0,\quad\pr_um\le0,
\]
which proves~\eqref{null:eq:bondi-mass}.

The scalar equation~\eqref{eq:ssesf} gives
\[
r\pr_u\pr_v\phi+\nu\pr_v\phi+\la\pr_u\phi=0,\qquad\pr_r(r\pr_u\phi)=-\nu\pr_r\phi.
\]
Hence
\begin{align*}
 r\pr_u\phi(u,r)&=-\int_0^r\nu(u,\rho)\pr_\rho\phi(u,\rho)d\rho,\\
 \int_R^\infty|\nu\pr_\rho\phi|d\rho&\les R^{-1/2}\|\rho\pr_\rho\phi\|_{L^2(d\rho)}
 \les R^{-1/2}\|\Phi-c\|_{L^2(1,\infty)}.
\end{align*}
Thus $r\pr_u\phi$ has a locally uniform, continuous limit. For $1\le a\le b$, integration of~\eqref{eq:ssesf} at fixed $V$ gives
\begin{align*}
 M_B(b)-M_B(a)=\lim_{V\to\infty}\int_a^b\frac{1-\mu(u,V)}{2\nu(u,V)}|r\pr_u\phi(u,V)|^2du=\int_a^b\frac1{2\nu_\infty(u)}
 \left|\lim_{v\to\infty}r\pr_u\phi(u,v)\right|^2du.
\end{align*}
By~\eqref{null:eq:metric-limits},
\[
U'=-2\nu_\infty,\qquad\frac{\Omega^2}{\la U'}\to2,\qquad\frac{\Omega^2\nu}{\la(U')^2}\to-1.
\]
Therefore $M_B\in C^1$ and
\[
-\frac{dM_B}{dU}=-\frac1{U'}\frac{dM_B}{du}=\frac1{4\nu_\infty(u)^2}\left|\lim_{v\to\infty}r\pr_u\phi(u,v)\right|^2.
\]
Integrating it by $U$ and combining with \eqref{null:eq:bondi-mass} give \eqref{null:eq:total-flux}.
If, in addition, $\Phi-c\in L^1$, then
\[
 |\phi-c|\le r^{-1/2}\|\pr_r(r\phi)-c\|_{L^2(dr)}\to0,\qquad
 \pr_u[r(\phi-c)]=r\pr_u\phi+\nu(\phi-c).
\]
At fixed $V$, integration from $1$ to $u$ gives
\[
r(\phi-c)(u,V)=\frac12\int_1^V(\Phi-c)dv+\int_1^u\pr_{u'}[r(\phi-c)](u',V)du'.
\]
For the second integral, $\nu(\phi-c)\to0$ locally uniformly in $u$, and we use the locally uniform limit of $r\pr_u\phi$. Thus
\[
\lim_{v\to\infty}r(\phi-c)(u,v)=\frac12\int_1^\infty(\Phi-c)dv+\int_1^u\lim_{v\to\infty}r\pr_{u'}\phi(u',v)du'.
\]
The integrand is continuous, so the limit is $C^1$ in $u$ with derivative $\lim_{v\to\infty}r\pr_u\phi$.
\end{proof}
\begin{proof}[Proof of Theorem~\ref{thm:null-infinity}\textup{(ii)}]
For $q\ge1$,
\[
 \int_1^\infty v^s|\Phi'(v)|^2dv
 =\|\Phi\|_{X^{s,1}}^2\le\|\Phi\|_{X^{s,q}}^2<\infty.
\]
For $s>1$, Cauchy--Schwarz gives $\Phi_\infty=\lim_{v\to\infty}\Phi(v)$ and
\[
 v^{s-1}|\Phi(v)-\Phi_\infty|^2
 \le\frac1{s-1}\int_v^\infty z^s|\Phi'(z)|^2dz\to0.
\]
Integration by parts and Cauchy--Schwarz give
\begin{align*}
 (s-1)\int_1^R v^{s-2}|\Phi-\Phi_\infty|^2dv\le R^{s-1}|\Phi(R)-\Phi_\infty|^2+2\left(\int_1^R v^{s-2}|\Phi-\Phi_\infty|^2dv\right)^{1/2}
 \left(\int_1^R v^s|\Phi'|^2dv\right)^{1/2}.
\end{align*}
Letting $R\to\infty$, we obtain
\[
 \int_1^\infty v^{s-2}|\Phi-\Phi_\infty|^2dv
 \le\frac4{(s-1)^2}\int_1^\infty v^s|\Phi'|^2dv.
\]
Thus $\Phi-\Phi_\infty\in L^2$ for $s\ge2$. For $s>3$,
\[
 \int_1^\infty|\Phi-\Phi_\infty|dv
 \le\left(\int_1^\infty v^{s-2}|\Phi-\Phi_\infty|^2dv\right)^{1/2}
 \left(\int_1^\infty v^{2-s}dv\right)^{1/2}<\infty.
\]
Proposition~\ref{null:prop:mass-radiation} gives the conclusions.
\end{proof}
\subsection{Regularity of the conformal metric}
\begin{proof}[Proof of Theorem~\ref{thm:null-infinity}\textup{(iii)}]
Fix $A>1$. For $1<s\le2$, the weighted estimate in \textup{(ii)},~\eqref{null:eq:profile-transport}, and radial averaging give
\begin{align*}
|\Phi(v)-\Phi_\infty|^2&\le\frac{v^{1-s}}{s-1}\int_v^\infty z^s|\Phi'(z)|^2dz=o(v^{1-s}),&|\pr_r(r\phi)(u,r)-\Phi_\infty|&=o_A(r^{-\frac{s-1}{2}}),\\
|\phi(u,r)-\Phi_\infty|&\le\frac1r\int_0^r|\pr_\rho(\rho\phi)-\Phi_\infty|d\rho=o_A(r^{-\frac{s-1}{2}}),&
r|\pr_r\phi|=|\pr_r(r\phi)-\phi|&=o_A(r^{-\frac{s-1}{2}}).
\end{align*}
For $1<s<2$, extend $\phi$ by $\Phi_\infty$ at $x=0$. For $0\le y\le x$ sufficiently small,
\begin{align*}
 |\pr_x\phi|&=r^2|\pr_r\phi|=o_A(x^{(s-3)/2}),\\
 |\phi(u,x)-\phi(u,y)|&\les_A\int_y^x z^{(s-3)/2}dz\les_A|x-y|^{(s-1)/2}.
\end{align*}
By~\eqref{half:eq:metric-log} and $\pr_r\nu=G$,
\begin{align*}
\pr_x\!\left(\frac{\Omega^2}{\la}\right)&=-\frac{\Omega^2}{\la}r^3|\pr_r\phi|^2=o_A(x^{s-2}),\\
\pr_x\!\left(x^2\frac{\Omega^2\nu}{\la}\right)&=x\frac{\Omega^2\nu}{\la}\left(2-\frac{\mu}{1-\mu}-r^2|\pr_r\phi|^2\right)=O_A(x).
\end{align*}
For $1<s<2$ and $0\le y\le x$ sufficiently small,
\[
 \left|\frac{\Omega^2}{\la}(u,x)-\frac{\Omega^2}{\la}(u,y)\right|
 \les_A\int_y^x z^{s-2}dz\les_A|x-y|^{s-1}.
\]
For $s-1<\beta\le1$, choose $(s-1)/2<\alpha<\beta/2$. The power data in \textup{(iv)(b)} belong to $X^{s,1}$ and satisfy
\[
 \frac{\Omega^2}{\la}(1,0)-\frac{\Omega^2}{\la}(1,x)\asymp x^{2\alpha},
\]
which excludes $\b$--H\"older regularity.

For $s\ge2$, take $s=2$ above. By~\eqref{eq:incoming-scalar-integral},
\[
 \pr_r\phi=o_A(r^{-3/2}),\qquad
 \pr_u\phi=-\frac1r\int_0^r\nu\pr_\rho\phi d\rho=O_A(r^{-1}),\qquad
 \pr_x(x^2\g)\to0.
\]
Since $\left.\pr_u\right|_x=\pr_u-\nu\pr_r$,
\[
 \left.\pr_u\right|_x\pr_r\phi
 =-\frac{\pr_u\phi}{r}
 -\left(\frac\nu r+G\right)\pr_r\phi-\nu\pr_r^2\phi.
\]
At fixed $R$, integration by parts yields
\begin{align*}
\frac{d}{du}\int_0^R\rho|\pr_\rho\phi|^2d\rho&=-2\int_0^R\pr_\rho\phi\,\pr_u\phi d\rho-2\int_0^R(\nu+\rho G)|\pr_\rho\phi|^2d\rho-2\int_0^R\rho\nu\pr_\rho\phi\,\pr_\rho^2\phi d\rho\\
&=-R\nu(u,R)|\pr_r\phi(u,R)|^2-\int_0^R(\nu+\rho G)|\pr_\rho\phi|^2d\rho-2\int_0^R\pr_\rho\phi\,\pr_u\phi d\rho.
\end{align*}
Uniformly for $1\le u\le A$,
\[
R|\nu(u,R)||\pr_r\phi(u,R)|^2+\int_R^\infty\left(|\nu+\rho G||\pr_\rho\phi|^2+2|\pr_\rho\phi\,\pr_u\phi|\right)d\rho
 \les_A R^{-3/2}\to0.
\]
Thus
\begin{align*}
 \frac{\Omega^2}{\la}(u,r)
 &=4\la(u,u)\exp\left(\int_0^r\rho|\pr_\rho\phi|^2d\rho\right),\qquad
 \nu_\infty\in C^1,\\
 \left.\pr_u\right|_x\frac{\Omega^2}{\la}(u,x)
 &\to-4\frac{d\nu_\infty}{du}(u).
\end{align*}
Finally,~\eqref{eq:ssesf} and~\eqref{eq:mass-radial} imply
\[
 \left.\pr_u\right|_x\mu
 =\frac{(1-\mu)r}{\nu}
 \left(|\pr_u\phi|^2-\nu^2|\pr_r\phi|^2\right)=O_A(r^{-1}),
\]
\[
x^2\frac{\Omega^2\nu}{\la}=-\frac{x^2}{4}(1-\mu)\left(\frac{\Omega^2}{\la}\right)^2,\qquad\left.\pr_u\right|_x\!\left(x^2\frac{\Omega^2\nu}{\la}\right)=O_A(x^2).
\]
Together with~\eqref{null:eq:conformal-metric}, these give the $C^1$ extension in the fixed $(u,x,\omega)$ coordinates.
\end{proof}
\begin{rk}
For $1<s<3$, Klainerman and the present authors~\cite[Definition~4.8 and Theorem~4.10]{KSW} obtain $C^{0,(s-1)/2}$ regularity of the conformal metric up to null infinity under $(s,q)$--asymptotic vacuum hypotheses. Without spherical symmetry, the null structure equations contain linear curvature terms~\cite[Appendix~A.1]{KSW}. In our case, the source term $r|\pr_r\phi|^2$ in~\eqref{half:eq:metric-log} is quadratic instead of linear. Together with the derivative estimates we obtained in the proof above, this yields $C^{0,s-1}$ regularity for the conformal metric, for $1<s<2$.
\end{rk}
\subsection{Examples at the thresholds}
\begin{proof}[Proof of Theorem~\ref{thm:null-infinity}\textup{(iv)}]
Fix $A>1$. We have
\[
r(u,v)=\frac{v-u}{2}+\int_1^u\left(\nu(u',v)+\frac12\right)du',\qquad\left|r(u,v)-\frac{v-u}{2}\right|\les\eps_0(u-1).
\]
Hence $r(u,v)=v/2+O_A(1)$ uniformly for $1\le u\le\min\{A,v\}$.\\ \\
\textup{(a)} For the oscillatory data $\Phi(v)=\sin\bigl(a\log(1+\log v)\bigr)$, $0<a\le\eps_0$ as in~\eqref{Phiosc}, direct computation shows $\|\Phi\|_{X^{1,1}}^2\le a^2\le\eps_0^2$. Integration in $v$ gives
\[
|\phi(1,v)-\Phi(v)|\le\frac1{v-1}\int_1^v(z-1)|\Phi'(z)|dz\le\frac{a}{v-1}\int_1^v\frac{dz}{1+\log z}\to0.
\]
For bounded $\Phi$,~\eqref{null:eq:profile-transport} and~\eqref{eq:lambda-u} give, uniformly for $1\le u\le A$,
\[
r\phi(u,v)=\frac12\int_1^v\Phi(z)dz+O_A(\log v),\qquad r(u,v)=\frac{v-1}{2}+O_A(1).
\]
Hence
\[
 \phi(u,v)-\phi(1,v)=O_A\!\left(\frac{\log v}{v}\right),\qquad
 \phi(u,v)=\Phi(v)+o_A(1).
\]
Therefore, such $\phi(u,v)$ fails to converge on every $C_u$ as $v\to\infty$ due to oscillation of $\Phi(v)$.\\ \\
\textup{(b)} Fix $1\le s<2$ and $(s-1)/2<\alpha<1/2$. Take
\begin{equation}\label{null:eq:power-data}
 \Phi(v)=\epsilon v^{-\alpha}.
\end{equation}
Then $\int_1^\infty|\Phi'(v)|dv=\epsilon$ and
\[
\|\Phi\|_{X^{s,1}}^2=\frac{\alpha^2\epsilon^2}{1+2\alpha-s},\qquad
\phi(1,r)=\frac{\epsilon\bigl((1+2r)^{1-\alpha}-1\bigr)}{2(1-\alpha)r}\sim\frac{\epsilon2^{-\alpha}}{1-\alpha}r^{-\alpha}.
\]
By~\eqref{eq:mass-radial} and $\mu\to0$,
\[
m(1,r)\sim\frac{\alpha^2\epsilon^22^{-2\alpha}}{2(1-\alpha)^2(1-2\alpha)}r^{1-2\alpha}\to\infty.
\]
Moreover,
\[
\Omega^2(1,r)=\exp\left(\int_0^r\rho|\pr_\rho\phi|^2d\rho\right),\qquad-2\nu_\infty(1)-\Omega^2(1,r)\sim\frac{-\alpha\nu_\infty(1)\epsilon^2 2^{-2\alpha}}{(1-\alpha)^2}r^{-2\alpha}.
\]
Since $\la(1,v)=1/2$,
\[
 \frac{\frac{\Omega^2}{\la}(1,0)-\frac{\Omega^2}{\la}(1,x)}x
 \asymp x^{2\alpha-1}\to\infty.
\]
For $1\le u\le A$,~\eqref{null:eq:profile-transport} and~\eqref{eq:lambda-u} give
\[
 \pr_r(r\phi)(u,v)=\epsilon v^{-\alpha}+O_A(v^{-1}),\qquad
 \la(u,v)=\frac12+O_A(v^{-1}),\qquad r(u,v)=\frac v2+O_A(1).
\]
Integration in $r$ yields
\[
 \phi(u,v)\sim\frac{\epsilon}{1-\alpha}v^{-\alpha},\qquad
 r\pr_r\phi(u,v)\sim-\frac{\alpha\epsilon}{1-\alpha}v^{-\alpha}.
\]
Hence
\[
 \pr_vm=\frac12(1-\mu)\la|r\pr_r\phi|^2
 \asymp_{\alpha,A}\epsilon^2v^{-2\alpha},\qquad M_B(u)=\infty.
\]

\noindent\textup{(c)} For $1/2<\alpha\le1$, choose smooth $\Phi$, vanishing near $v=1$, with $|\Phi|+|\Phi'|\les_\alpha\epsilon$ on $[1,2]$ and
\begin{equation}\label{null:eq:logarithmic-data}
 \Phi(v)=\frac{\epsilon}{v(\log v)^\alpha}\qquad(v\ge2).
\end{equation}
Then
\begin{align*}
 \int_1^\infty|\Phi'(v)|dv
 &=\int_1^2|\Phi'(v)|dv
 +\epsilon\int_2^\infty\frac{1+\alpha/\log v}{v^2(\log v)^\alpha}dv\\
 &=\int_1^2|\Phi'(v)|dv
 +\frac{\epsilon}{2(\log2)^\alpha}\les_\alpha\epsilon.
\end{align*}
Moreover,
\[
\int_1^\infty v^3|\Phi'(v)|^2dv\les_\alpha\epsilon^2,\qquad\Phi\in L^2(1,\infty),\qquad\int_1^\infty\Phi(v)dv=\infty.
\]
Proposition~\ref{null:prop:mass-radiation} gives finite mass, $M_B(u)\downarrow0$, and~\eqref{null:eq:total-flux}. Moreover,
\begin{align*}
 r\phi(u,V)=\frac12\int_1^V\Phi(v)dv
 +\int_1^u\bigl[r\pr_{u'}\phi+\nu\phi\bigr](u',V)du'=\frac12\int_1^V\Phi(v)dv+O_A(1)\to\infty.
\end{align*}
This concludes the proof of Theorem~\ref{thm:null-infinity}\textup{(iv)}.
\end{proof}
\begin{rk}\label{null:rk:earlier-existence}
The power \eqref{null:eq:power-data} and logarithmic \eqref{null:eq:logarithmic-data} examples above also satisfy the small-total-variation criterion of~\cite{ChristodoulouBV} and the hypotheses of~\cite[Theorem~1.1]{LOY}.
\end{rk}

\begin{rk}\label{null:rk:mass-curvature}
Take nonzero $\Phi\in C_c^\infty((2,3))$ with small $X^{1,1/2}$ seminorm. By~\eqref{eq:mass-radial} and the scaling in Section~\ref{sec:scaling-criticality},
\[
 0<M_B(1)<\infty,\qquad
 \|\Phi_A\|_{X^{1,1/2}}=\|\Phi\|_{X^{1,1/2}},\qquad
 \lim_{v\to\infty}m_A(1,v)=A M_B(1).
\]
The spacetime sectional curvature of an angular tangent plane is
\[
\frac{1-\g(\nabla r,\nabla r)}{r^2}=\frac{1+4\Omega^{-2}\la\nu}{r^2}=\frac{2m}{r^3},\qquad\quad\frac{2m_A}{r_A^3}(u,v)=A^{-2}\frac{2m}{r^3}(\tilde u,\tilde v).
\]
As $A\to\infty$, the initial Bondi mass is arbitrarily large; as $A\downarrow0$, the curvature is arbitrarily large. The critical seminorm and future completeness are preserved.
\end{rk}
\appendix
\section{Higher regularity and continuation}\label{app:continuation}
\begin{prop}\label{app:prop:continuation}
Let $s\geq 1$ and $q\in\{1,\tfrac12\}$. Suppose that the regular solution of \eqref{eq:ssesf}--\eqref{eq:data} satisfies the improved bootstrap bounds of Theorem~\ref{thm:s=1} for $q=1$, or of Theorem~\ref{thm:half-stability} for $q=\tfrac12$, on $1\le u<u^*<\infty$. If $\|\Phi\|_{C^k(C_1)}<\infty$ for some integer $k\ge1$, then
\[
\la,\quad\nu,\quad\mu,\quad\Om^2,\quad G,\quad\phi,\quad\pr_r(r\phi),\quad\pr_v(r\phi),\quad\pr_u(r\phi)
\]
have bounded $C^k$ norms up to $C_{u^*}$, and the solution extends beyond $C_{u^*}$ as a regular $C^k$ solution. For smooth data the extension is smooth. On $v\le V$, with $V>u^*$, it suffices to assume $\|\Phi\|_{C^k([1,V])}<\infty$.
\end{prop}
\begin{proof}
Fix $1<U<u^*$ and $V>U$. Recall that $\mathcal T_V=\{(u,v):1\le u\le v\le V\}$. We work on $\mathcal T_V\cap\{1\leq u\leq U\}$. In both cases $\la,-\nu,1-\mu\asymp1$ and $r\asymp v-u$. All estimates below are uniform in $U$. They are uniform in $V$ when the data norm is taken on $C_1$. We proceed hierarchically by differentiation order and region.

\noindent\textbf{Step 1. First derivative estimates.} We begin with the zero- and first-order bounds on $\mathcal T_V\cap\{1\le u\le U\}$.
For $q=1$, \eqref{prr2rphi} and Proposition~\ref{prop:mass-mu-G} give
\begin{align*}
|\pr_r^2(r\phi)(u,v)|&\les |\Phi'(v)|+\frac{\eps^3}{vL(v)^3},\\
\int_1^u\|G(u',\cdot)\|_{L^\infty(u',V)}du'&\les\eps^2\int_1^u\frac{du'}{u'L(u')^2}\les\eps^2,
\end{align*}
where we used \eqref{L-comparison} and \eqref{borderline-property}. For $q=\tfrac12$, Step~2 of Section~\ref{half:sec:proof}, \eqref{half:eq:G-sup-bound} and \eqref{half:eq:closed-energy} give
\begin{align*}
\|\la^2\pr_r^2(r\phi)\|_{L^\infty(u,V)}&\le\tfrac12\|\Phi'\|_{L^\infty(1,V)}e^{CB_V(u)},\\
\int_1^u\|G(u',\cdot)\|_{L^\infty(u',V)}du'&\les B_V(u)\le2\eps_0^2.
\end{align*}
On each $C_u$, integration in $r$ from the axis gives
\begin{equation}\label{app:average}
\phi(u,r)=\frac1r\int_0^r\pr_{r'}(r'\phi)(u,r')dr',\qquad\pr_r^i\phi(u,r)=\int_0^1t^i\pr_r^{i+1}(r\phi)(u,tr)dt.
\end{equation}
Using \eqref{half:eq:h-transport} and \eqref{eq:data}, integrate at fixed $v$, take the supremum and apply Gr\"onwall to get
\begin{align*}
\|\pr_r(r\phi)\|_{L^\infty(u,V)}&\le\|\Phi\|_{L^\infty(1,V)}+2\int_1^u\|G\|_{L^\infty(u',V)}\|\pr_r(r\phi)\|_{L^\infty(u',V)}du'\\
&\le\|\Phi\|_{L^\infty(1,V)}\exp\!\left(2\int_1^u\|G\|_{L^\infty(u',V)}du'\right).
\end{align*}
For the geometry, \eqref{half:eq:G-radial} and \eqref{half:eq:metric-log} imply
\begin{align}
 \frac{G(u,r)}{r}&=\int_0^1 t^2\left[\frac{\nu}{1-\mu}(\pr_r\phi)^2\right](u,tr)dt,
 &\pr_rG&=\frac{\nu}{1-\mu}(\pr_r\phi)^2-2\frac Gr,\label{app:radial-G}\\
 \pr_r\!\left(\frac{\nu}{1-\mu}\right)
 &=\frac{\nu}{1-\mu}\bigl(\pr_r(r\phi)-\phi\bigr)\pr_r\phi,
 &\pr_v\log\la(u,v)&=\int_1^u(\la\pr_rG)(u',v)du'.\label{app:radial-metric}
\end{align}
The last equality follows from \eqref{eq:lambda-u} and $\la(1,v)=1/2$. In particular, \eqref{app:average}, \eqref{app:radial-G} and \eqref{eq:G-definition} give
\[
|\pr_r\phi|\le\tfrac12\|\pr_r^2(r\phi)\|_{L^\infty(u,V)},\qquad\frac{|G|}{r}\les\min\!\left\{\|\pr_r\phi\|_{L^\infty(u,V)}^2,\frac1{r^2}\right\}\les\frac1{(1+r)^2}.
\]
This establishes the bounds on $\phi$, $\pr_r\phi$, $\pr_r(r\phi)$, $\pr_r^2(r\phi)$, and $|G|/r$.

\noindent\textbf{Step 2. Higher radial derivatives.} We next induct on $j=2,\dots,k$ to bound the $\pr_r$ and $\pr_v$ derivatives through order $k$.
Suppose that $\la,\nu/(1-\mu),G,\phi,\pr_r(r\phi)$ and their $v$ derivatives through order $j-1$ are bounded, where $2\le j\le k$. Implicit constants in this step may depend on these bounds. Since $\pr_v=\la\pr_r$ and $\pr_v\nu=\la G$, we also control $\pr_v^i\pr_r\phi$ for $i\le j-2$ and $\pr_v^i\nu$ for $i\le j-1$. For $0\le i\le j-1$, differentiating $r\pr_r\phi=\pr_r(r\phi)-\phi$ and $rG=\nu/(1-\mu)-\nu$ gives
\begin{align*}
 r\pr_v^i\pr_r\phi
 &=\pr_v^i\bigl(\pr_r(r\phi)-\phi\bigr)
 -\sum_{\ell=1}^i\binom i\ell (\pr_v^{\ell-1}\la)\pr_v^{i-\ell}\pr_r\phi,\\
 r\pr_v^i G
 &=\pr_v^i\!\left(\frac{\nu}{1-\mu}-\nu\right)
 -\sum_{\ell=1}^i\binom i\ell (\pr_v^{\ell-1}\la)\pr_v^{i-\ell}G.
\end{align*}
Consequently,
\begin{equation}\label{app:lower-products}
 \sum_{i=0}^{j-2}(1+r)|\pr_v^i\pr_r\phi|
 +\sum_{i=0}^{j-1}(1+r)|\pr_v^iG|
 +|r\pr_v^{j-1}\pr_r\phi|\les1 .
\end{equation}
We estimate the two terms in \eqref{eq:pruprrrphi} separately. By Leibniz and \eqref{app:lower-products},
\begin{align*}
 |\pr_v^{j-1}(\la^2G\pr_r\phi)|
 &\les \frac{|G|}{r}\,|r\pr_v^{j-1}\pr_r\phi|
 +\sum_{\ell=1}^{j-1}|\pr_v^\ell(\la^2G)|\,|\pr_v^{j-1-\ell}\pr_r\phi|
 \les (1+r)^{-2}.
\end{align*}
In the cubic term, all $j-1$ derivatives may fall on one factor $\pr_r\phi$. Every other term contains three factors with at most $j-2$ derivatives on each, and $|\pr_v^\ell(\la^2\nu r/(1-\mu))|\les 1+r$ for $0\le\ell\le j-1$. Hence
\begin{align*}
 \left|\pr_v^{j-1}\!\left[\la^2\frac{\nu}{1-\mu}r(\pr_r\phi)^3\right]\right|
 &\les |\pr_r\phi|^2\,|r\pr_v^{j-1}\pr_r\phi|
 +(1+r)\left(\sum_{i=0}^{j-2}|\pr_v^i\pr_r\phi|\right)^3
 \les (1+r)^{-2}.
\end{align*}
On $C_1$, $\la^2\pr_r^2(r\phi)=\Phi'/2$. First, to bound $\pr_v^j\pr_r(r\phi)$, we differentiate the integral of \eqref{eq:pruprrrphi} at fixed $v$ and use $-\pr_ur=-\nu\asymp1$ to get
\begin{align}
\begin{split}\label{app:commuted}
 \left|\pr_v^{j-1}\!\left(\la^2\pr_r^2(r\phi)\right)(u,v)-\tfrac12\Phi^{(j)}(v)\right|&\le\int_1^u\left|\pr_v^{j-1}\!\left[3\la^2G\pr_r\phi
 -\la^2\frac{\nu}{1-\mu}r(\pr_r\phi)^3\right](u',v)\right|du'\\
 &\les\int_{r(u,v)}^{r(1,v)}\frac{dr'}{(1+r')^2}
 =\frac{r(1,v)-r(u,v)}{(1+r(u,v))(1+r(1,v))}\\
 &\les\frac{u-1}{v(1+r(u,v))}.
 \end{split}
\end{align}
Since $\la^2\pr_r^2(r\phi)=\la\pr_v\pr_r(r\phi)$,
\[
 \la\pr_v^j\pr_r(r\phi)=\pr_v^{j-1}\!\left(\la^2\pr_r^2(r\phi)\right)
 -\sum_{\ell=1}^{j-1}\binom{j-1}{\ell}(\pr_v^\ell\la)\pr_v^{j-\ell}\pr_r(r\phi).
\]
Thus \eqref{app:commuted} yields $|\pr_v^j\pr_r(r\phi)(u,v)|\les1+|\Phi^{(j)}(v)|$. Next, for the $j$-th $\pr_r$-derivatives, \eqref{app:average}--\eqref{app:radial-metric} give
\begin{align*}
 |\pr_r^jG|&\le\left(1+\frac{2}{j+2}\right)
 \left\|\pr_r^{j-1}\!\left[\frac{\nu}{1-\mu}(\pr_r\phi)^2\right]\right\|_{L^\infty(u,V)},\\
 |\pr_r^jG|+\left|\pr_r^j\frac{\nu}{1-\mu}\right|+|\pr_r^j\phi|
 &\les 1+\|\pr_r^j\phi\|_{L^\infty(u,V)}
 \les 1+\|\pr_r^{j+1}(r\phi)\|_{L^\infty(u,V)}
 \les 1+\|\Phi^{(j)}\|_{L^\infty(1,V)}.
\end{align*}
In the last inequality use $\pr_r=\la^{-1}\pr_v$, which differentiates $\la$ at most $j-1$ times. Finally, for $\pr_v^j\log\la$,
\[
 |\pr_v^j\log\la(u,v)|\le\int_1^u|\pr_v^jG(u',v)|du'
 \les (u-1)\bigl(1+\|\Phi^{(j)}\|_{L^\infty(1,V)}\bigr).
\]
This closes the induction. Hence all $\pr_v^{\le k}$ and $\pr_r^{\le k}$ derivatives of $\la,\nu/(1-\mu),G,\phi,\pr_r(r\phi)$, together with $\pr_r^{k+1}(r\phi)$, are bounded.

\noindent\textbf{Step 3. Mixed derivatives near the axis.}
We first work on $v\le 2u^*+2$, where $r$ is bounded. The derivative in $u$ at fixed $r$ is $\pr_u-\nu\pr_r$. By $[\pr_u,\pr_r]=-G\pr_r$ and $\pr_r\nu=G$, it commutes with $\pr_r$. We induct on the number $i$ of derivatives at fixed $r$, keeping $i+j\le k$. Step~2 gives the case $i=0$ for the outgoing data and coefficients. For $i\ge1$, we first control the commuted $\pr_r$-derivatives and $\log\la$: \eqref{half:eq:h-transport} and \eqref{eq:lambda-u} give
\begin{align*}
 (\pr_u-\nu\pr_r)^i\pr_r^{j+1}(r\phi)
 &=-(\pr_u-\nu\pr_r)^{i-1}\pr_r^j
 \bigl[G(\pr_r(r\phi)-\phi)+\nu\pr_r^2(r\phi)\bigr],\\
 (\pr_u-\nu\pr_r)^i\pr_r^j\log\la
 &=(\pr_u-\nu\pr_r)^{i-1}\pr_r^j(G-\nu\pr_r\log\la).
\end{align*}
Both right hand sides are bounded by the induction hypothesis. Applying \eqref{app:average} at fixed $r$ gives
\[
 (\pr_u-\nu\pr_r)^i\pr_r^j\phi(u,r)
 =\int_0^1t^j\bigl[(\pr_u-\nu\pr_r)^i\pr_r^{j+1}(r\phi)\bigr](u,tr)dt .
\]
Next, we estimate $(\pr_u-\nu\pr_r)^i\pr_r^j[r(\pr_r\phi)^2]$. When all derivatives fall on one factor $\pr_r\phi$, use
\begin{align*}
r(\pr_u-\nu\pr_r)^i\pr_r^{j+1}\phi=(\pr_u-\nu\pr_r)^i\pr_r^{j+1}(r\phi)-(j+1)(\pr_u-\nu\pr_r)^i\pr_r^j\phi .
\end{align*}
The right hand side is bounded and vanishes at $r=0$ by \eqref{app:average}. The other terms in the product rule are bounded by
\[
 (1+r)\Bigg(\sum_{\substack{0\le\ell\le i,\ p\ge0\\ \ell+p\le k-1}}
 |(\pr_u-\nu\pr_r)^\ell\pr_r^{p+1}\phi|\Bigg)^2\les1 .
\]
Integrating from the axis then yields the bounds for $\nu/(1-\mu)$, $\nu$, $G$, and $\pr_u(r\phi)$: integrate \eqref{half:eq:metric-log}, $\pr_r(r\nu)=\nu/(1-\mu)$ and $\pr_r\pr_u(r\phi)=G\phi$ from the axis. Together with \eqref{app:radial-G}, this gives
\begin{align}
 \frac{\nu}{1-\mu}(u,r)&=-\la(u,0)\exp\!\left(\int_0^r r'|\pr_{r'}\phi(u,r')|^2dr'\right),
 &\nu(u,r)&=\int_0^1\frac{\nu}{1-\mu}(u,tr)dt,\label{app:metric-axis}\\
 G(u,r)&=\int_0^1 t\left[\frac{\nu}{1-\mu}r(\pr_r\phi)^2\right](u,tr)dt,\notag\\
 \pr_u(r\phi)(u,r)&=-\la(u,0)\phi(u,0)+\int_0^r(G\phi)(u,r')dr'.\label{app:scalar-axis}
\end{align}
Differentiating the exponential at fixed $r$ bounds $(\pr_u-\nu\pr_r)^i[\nu/(1-\mu)]$. For $j\ge1$ use
\[
(\pr_u-\nu\pr_r)^i\pr_r^j\!\left(\frac{\nu}{1-\mu}\right)=(\pr_u-\nu\pr_r)^i\pr_r^{j-1}\!\left[\frac{\nu}{1-\mu}r(\pr_r\phi)^2\right]
\]
successively in $j$. The integrals in \eqref{app:metric-axis}--\eqref{app:scalar-axis} give the same bounds for $\nu,G,\pr_u(r\phi)$. Finally, for $\mu$, $\Om^2$, and the $(u,v)$-derivatives, use
\[
 \mu=1-\frac{\nu}{\nu/(1-\mu)},\qquad \Om^2=-\frac{4\la\nu}{1-\mu},\qquad
 \pr_u=(\pr_u-\nu\pr_r)+\nu\pr_r,\qquad \pr_v=\la\pr_r .
\]
This completes all $C^k$ bounds on $v\le 2u^*+2$.

\noindent\textbf{Step 4. Mixed derivatives on the complete cones.} On $v\ge 2u^*+2$, we bound $\pr_u(r\phi)$, then the pure $\pr_u^i$ derivatives, and finally the mixed $\pr_u^i\pr_v^j$ derivatives.
For $v\ge 2u^*+2$ we have $r\asymp v$. Since $G\le0$, \eqref{eq:nu-radial} and \eqref{app:scalar-axis} give
\begin{align*}
\int_0^{r(u,v)}|G(u,r')|dr'&=\nu(u,0)-\nu(u,v)\les1,\\
|\pr_u(r\phi)(u,v)|&\le|\la\phi|(u,0)+\|\phi\|_{L^\infty(u,V)}\int_0^{r(u,v)}|G(u,r')|dr'\les1 .
\end{align*}
Next, for the pure $\pr_u^i$ derivatives, assume the pure $u$ derivatives through order $i-1$ are bounded, $1\le i\le k$. At fixed $v$, \eqref{eq:ssesf}, \eqref{eq:lambda-u} and \eqref{half:eq:h-transport} give
\begin{align*}
 r\pr_u\phi&=\pr_u(r\phi)-\nu\phi,
 & r\pr_u\mu&=\frac{1-\mu}{\nu}\bigl(\pr_u(r\phi)-\nu\phi\bigr)^2-\nu\mu,\\
 r\pr_u\log\la&=\frac{\nu\mu}{1-\mu},
 & r\pr_u\pr_r(r\phi)&=-\frac{\nu\mu}{1-\mu}\bigl(\pr_r(r\phi)-\phi\bigr).
\end{align*}
Apply $\pr_u^{i-1}$ and use $\pr_u^\ell r=\pr_u^{\ell-1}\nu$ for $\ell\ge1$. We obtain
\begin{equation}\label{app:exterior-u}
 |\pr_u^i\phi|+|\pr_u^i\mu|+|\pr_u^i\la|+|\pr_u^i\pr_r(r\phi)|\les r^{-1}.
\end{equation}
In particular, for $1\le\ell\le i$, $|\pr_u^\ell(\mu/(1-\mu))|\les r^{-1}$ and $|\pr_u^\ell(r^{-1})|\les r^{-2}$. Hence
\[
 \left|\pr_u^i\!\left(\frac G\nu\right)\right|
 \le\frac1r\left|\pr_u^i\frac{\mu}{1-\mu}\right|
 +\sum_{\ell=1}^i\binom i\ell
 |\pr_u^\ell(r^{-1})|\left|\pr_u^{i-\ell}\frac{\mu}{1-\mu}\right|\les r^{-2}.
\]
For $\nu$ and $\pr_u(r\phi)$ use $\pr_v\log(-\nu)=\la G/\nu$ and $\pr_v\pr_u(r\phi)=\la G\phi$. By \eqref{app:exterior-u},
\begin{align*}
 |\pr_v\pr_u^i\log(-\nu)|
 &=|\pr_u^i(\la G/\nu)|\les r^{-2},\\
 \left|\pr_v\pr_u^{i+1}(r\phi)-\la\phi\frac G\nu\pr_u^i\nu\right|
 &\le\sum_{\ell=1}^i\binom i\ell
 \left|\pr_u^\ell\!\left(\la\phi\frac G\nu\right)\right|
 |\pr_u^{i-\ell}\nu|\les r^{-2}.
\end{align*}
At $i=k$ one differentiates the equations integrated in $v$. Using the bounds at $v=2u^*+2$ from Step~3, integrate the first equation and then the second to get
\begin{align*}
 |\pr_u^i\nu(u,v)|&\les1+\int_{2u^*+2}^v\frac{dv'}{(v')^2}\les1,\\
 |\pr_u^{i+1}(r\phi)(u,v)|
 &\les1+\int_{2u^*+2}^v\!\left(\frac1{(v')^2}+|G|\,|\pr_u^i\nu|\right)dv'
 \les1+\int_0^{r(u,v)}|G(u,r')|dr'\les1 .
\end{align*}
This closes the induction in $i$. Finally, for the mixed derivatives $\pr_u^i\pr_v^j$ with $i,j\ge1$ and $i+j\le k$, use Step~2 and
\[
 \pr_v\nu=\la G,\qquad \pr_v\pr_u(r\phi)=\la G\phi .
\]
Let $i,j\ge1$, $i+j\le k$. Assuming the bounds through order $i+j-1$, we have $r\asymp v$ and
\begin{align*}
 |\pr_u^i\pr_v^j\phi|
 &=\left|\pr_u^{i-1}\pr_v^j\left[\frac{\pr_u(r\phi)-\nu\phi}{r}\right]\right|\les r^{-1},\\
 |\pr_u^i\pr_v^j\mu|
 &=\left|\pr_u^{i-1}\pr_v^j\left[\frac{1-\mu}{r\nu}\bigl(\pr_u(r\phi)-\nu\phi\bigr)^2-\frac{\nu\mu}{r}\right]\right|\les r^{-1},\\
 |\pr_u^i\pr_v^j\la|+|\pr_u^i\pr_v^j\pr_r(r\phi)|
 &=|\pr_u^{i-1}\pr_v^j(\la G)|
 +\left|\pr_u^{i-1}\pr_v^j\bigl[G(\pr_r(r\phi)-\phi)\bigr]\right|\les r^{-1},\\
 |\pr_u^i\pr_v^j\nu|+|\pr_u^i\pr_v^j\pr_u(r\phi)|
 &=|\pr_u^i\pr_v^{j-1}(\la G)|+|\pr_u^i\pr_v^{j-1}(\la G\phi)|\les r^{-1}.
\end{align*}
Here $G=\nu\mu/[r(1-\mu)]$ by \eqref{eq:G-definition}. The bounds for $\Om^2$ and $\pr_v(r\phi)$ follow from $\Om^2=-4\la\nu/(1-\mu)$ and $\pr_v(r\phi)=\la\pr_r(r\phi)$. Together with Steps~1--3, this establishes all $C^k$ bounds in the statement.

\noindent\textbf{Step 5. Continuation of the solution.}
Fix $u^*<V<\infty$ and $U<u^*$ sufficiently close to $u^*$. Introduce the normalized coordinates
\[
\widehat v=1+2r(U,v),\qquad\quad \widehat u=1+2r(U,u).
\]
The cone $C_U$ is given by $\widehat u=1$, and
$\pr_{\widehat v}r=1/2$ there. Moreover,
\[
\pr_{\widehat v}^j
\bigl(2\pr_{\widehat v}(r\phi)\bigr)
=2^{-j}\pr_r^{j+1}(r\phi),
\qquad 0\leq j\leq k.
\]
By Steps~1--2, the normalized characteristic data on
$C_U\cap\{v\le V\}$ have uniformly bounded $C^k$ norms,
independently of $U$. Extend these data beyond
$\widehat v(V)$ with uniformly bounded $C^k$ norms.
The local characteristic existence theorem
of~\cite[Section~2.1]{LOY} then provides a lifespan
independent of $U$. Since
\[
\widehat u(u^*)-1
=2r(U,u^*)
=2\int_U^{u^*}\la(U,v')\,dv'
\les u^*-U\to0,
\]
the restarted solution extends beyond $C_{u^*}$
for $U$ sufficiently close to $u^*$ and agrees
with the original solution by uniqueness.
When $\|\Phi\|_{C^k(C_1)}<\infty$, the same argument
applies on the entire outgoing cone, using the
bounds uniform in $V$. For smooth data, the argument
applies at every order $k\ge1$, yielding a smooth
continuation.

Finally, to conclude global existence, fix $V>1$ and let $u^*\le V$ be the maximal bootstrap time on $\mathcal T_V$. The bootstrap improvements in the proofs of Theorems \ref{thm:s=1} and \ref{thm:half-stability}, together with the preceding continuation argument, imply $u^*=V$. Taking a sequence $V_n\to\infty$, uniqueness on overlapping triangles gives a unique global smooth solution on $1\le u\le v<\infty$.
\end{proof}

\end{document}